\documentclass[twocolumn]{./aastex63}

\usepackage{amsmath}
\usepackage{empheq}
\usepackage{mathrsfs}
\usepackage{textcomp}
\usepackage{enumitem}   
\usepackage{gensymb}
\usepackage{hyperref}
\usepackage{graphicx}
\usepackage[caption=false]{subfig}
\usepackage{multirow}
\usepackage{longtable}

\usepackage{lineno}
\definecolor{DarkOrange}{RGB}{204, 85, 0}
\definecolor{LincolnGreen}{RGB}{17, 102, 0}

\newcommand{\rztf}{$r_\mathrm{ZTF}$}
\newcommand{\gztf}{$g_\mathrm{ZTF}$}
\newcommand{\tfl}{$t_\mathrm{fl}$}
\newcommand{\trise}{$t_\mathrm{rise}$}
\newcommand{\tbmax}{$T_{B,\mathrm{max}}$}
\received{2020 January 17}
\revised{2020 August 4}
\accepted{2020 August 6}
\submitjournal{ApJ}

\shorttitle{The Rise of ZTF SNe\,Ia}
\shortauthors{Miller et al.}
\watermark{DRAFT}
\begin{document}

\title{ZTF Early Observations of Type Ia  Supernovae II: \\ First Light, the Initial Rise, and Time to Reach Maximum Brightness}

\author[0000-0001-9515-478X]{A.~A.~Miller}
\affiliation{Center for Interdisciplinary Exploration and Research in Astrophysics and Department of Physics and Astronomy, Northwestern University, 1800 Sherman Ave, Evanston, IL 60201, USA}
\affiliation{The Adler Planetarium, Chicago, IL 60605, USA}
\email{amiller@northwestern.edu}

\author[0000-0001-6747-8509]{Y.~Yao}
\affiliation{Cahill Center for Astrophysics, 
             California Institute of Technology, 
             1200 E.~California Boulevard, Pasadena, CA 91125, USA}

\author[0000-0002-8255-5127]{M.~Bulla} 
\affiliation{Nordita, KTH Royal Institute of Technology and Stockholm
University, Roslagstullsbacken 23, SE-106 91 Stockholm, Sweden}

\author[0000-0002-1128-3662]{C.~Pankow}
\affiliation{Center for Interdisciplinary Exploration and Research in Astrophysics and Department of Physics and Astronomy, Northwestern University, 1800 Sherman Ave, Evanston, IL 60201, USA}

\author[0000-0001-8018-5348]{E.~C.~Bellm}
\affiliation{DIRAC Institute, Department of Astronomy, University of Washington, 3910 15th Avenue NE, Seattle, WA 98195, USA}

\author[0000-0003-1673-970X]{S.~B.~Cenko}
\affiliation{Astrophysics Science Division, NASA Goddard Space Flight Center, 8800 Greenbelt Road, Greenbelt, MD 20771, USA}
\affiliation{Joint Space-Science Institute, University of Maryland, College Park, MD 20742, USA}

\author{R.~Dekany}
\affiliation{Caltech Optical Observatories, California Institute of Technology, Pasadena, CA 91125, USA}
\author[0000-0002-4223-103X]{C.~Fremling}
\affiliation{Cahill Center for Astrophysics, 
             California Institute of Technology, 
             1200 E.~California Boulevard, Pasadena, CA 91125, USA}

\author[0000-0002-3168-0139]{M.~J.~Graham}
\affiliation{Division of Physics, Mathematics, and Astronomy,
             California Institute of Technology, Pasadena, CA
             91125, USA}

\author[0000-0002-6540-1484]{T.~Kupfer}
\affiliation{Kavli Institute for Theoretical Physics, University of California, Santa Barbara, CA 93106, USA}
\author[0000-0003-2451-5482]{R.~R.~Laher}
\affiliation{IPAC, California Institute of Technology, 1200 E.~California
             Blvd, Pasadena, CA 91125, USA}

\author[0000-0003-2242-0244]{A.~A.~Mahabal}
\affiliation{Division of Physics, Mathematics, and Astronomy,
             California Institute of Technology, Pasadena, CA
             91125, USA}
\affiliation{Center for Data Driven Discovery, California Institute of Technology, Pasadena, CA 91125, USA}

\author[0000-0002-8532-9395]{F.~J.~Masci}
\affiliation{IPAC, California Institute of Technology, 1200 E.~California
             Blvd, Pasadena, CA 91125, USA}

\author[0000-0002-3389-0586]{P.~E.~Nugent}
\affiliation{Computational Cosmology Center, Lawrence Berkeley National Laboratory, 1 Cyclotron Road, Berkeley, CA 94720, USA}
\affiliation{Department of Astronomy, University of California, Berkeley, CA 94720-3411, USA}

\author[0000-0002-0387-370X]{R.~Riddle}
\affiliation{Caltech Optical Observatories, California Institute of Technology, Pasadena, CA 91125, USA}

\author[0000-0001-7648-4142]{B.~Rusholme}
\affiliation{IPAC, California Institute of Technology, 1200 E.~California
             Blvd, Pasadena, CA 91125, USA}

\author[0000-0001-7062-9726]{R.~M.~Smith}
\affiliation{Caltech Optical Observatories, California Institute of Technology, Pasadena, CA 91125, USA}

\author[0000-0003-4401-0430]{D.~L.~Shupe}
\affiliation{IPAC, California Institute of Technology, 1200 E.~California
             Blvd, Pasadena, CA 91125, USA}

\author[0000-0002-2626-2872]{J.~van Roestel}
\affiliation{Division of Physics, Mathematics, and Astronomy,
             California Institute of Technology, Pasadena, CA
             91125, USA}

\author[0000-0001-5390-8563]{S.~R.~Kulkarni}
\affiliation{Cahill Center for Astrophysics, 
             California Institute of Technology, 
             1200 E.~California Boulevard, Pasadena, CA 91125, USA}



\begin{abstract}

While it is clear that Type Ia supernovae (SNe) are the result of
thermonuclear explosions in C/O white dwarfs (WDs), a great deal remains
uncertain about the binary companion that facilitates the explosive disruption
of the WD. Here, we present a comprehensive analysis of a large, unique data
set of 127 SNe\,Ia with exquisite coverage by the Zwicky Transient Facility
(ZTF). High-cadence (six observations per night) ZTF observations allow us to
measure the SN rise time and examine its initial evolution. We develop a
Bayesian framework to model the early rise as a power law in time, which
enables the inclusion of priors in our model. For a volume-limited subset of
normal SNe\,Ia, we find that the mean power-law index is consistent with 2 in
the \rztf-band ($\alpha_r = 2.01\pm0.02$), as expected in the expanding
fireball model. There are, however, individual SNe that are clearly
inconsistent with $\alpha_r=2$. We estimate a mean rise time of 18.9\,d (with
a range extending from $\sim$15 to 22\,d), though this is subject to the
adopted prior. We identify an important, previously unknown, bias whereby the
rise times for higher-redshift SNe within a flux-limited survey are
systematically underestimated. This effect can be partially alleviated if the
power-law index is fixed to $\alpha=2$, in which case we estimate a mean rise
time of 21.7\,d (with a range from $\sim$18 to 23\,d). The sample includes a
handful of rare and peculiar SNe\,Ia. Finally, we conclude with a discussion
of lessons learned from the ZTF sample that can eventually be applied to
observations from the Vera C.~Rubin Observatory.

\end{abstract}


\vspace{1em}

\section{Introduction}\label{sec:intro}

The fact that supernovae (SNe) of Type Ia can be empirically calibrated as
standardizable candles has made them arguably the most important tool in all of
physics for the past $\sim$two decades. By unlocking our ability to accurately
measure distances at high redshift, SNe\,Ia have revolutionized our
understanding of the universe \citep{Riess98,Perlmutter99}. 

While it is all but certain that SNe\,Ia are the result of thermonuclear
explosions in carbon-oxygen (C/O) white dwarfs (WDs) in binary star systems
(see \citealt{Maoz14,Livio18}), there remains a great deal about SNe\,Ia
progenitors and the precise explosion mechanism that we do not know. This
leads to the tantalizing hope that an improved understanding of the binary
companions or explosion may improve our ability to calibrate these
standardizable candles.

Pinning down the binary companion to the exploding WD remains particularly
difficult. There are likely two dominant pathways toward explosion. In the
first, the WD accretes H or He from, or merges with the core of, a
nondegenerate companion and eventually explodes as it approaches the
Chandrasekhar mass ($M_\mathrm{Ch}$; known as the single degenerate, or SD,
scenario \citet{Whelan73}). In the second scenario, the explosion follows the
interaction or merger of two WD stars (known as the double degenerate, or DD,
scenario \citet{Webbink84}). While the debate has long focused on which of
these two scenarios is correct, empirical evidence supports both
channels. PTF\,11kx, an extreme example of a SN Ia, showed evidence of
multiple shells of H-rich circumstellar material \citep{Dilday12},
which is precisely what one would expect in a WD$+$red giant system that had
undergone multiple novae prior to the final, fatal thermonuclear explosion
(see \citet{Soker13} for an alternative explanation for PTF\,11kx). On the
other hand, hypervelocity WDs discovered by Gaia are likely the surviving
companions of DD explosions \citep{Shen18}.

Recent observational evidence also challenges the canonical picture
that WDs only explode near, or in excess of, $M_\mathrm{Ch}$. Detailed
modeling of SNe\,Ia light curves \citep{Scalzo14a} and a blue-to-red-to-blue
color evolution observed in a few SNe \citep{Jiang17,De19,Bulla20}, point to
sub-$M_\mathrm{Ch}$ mass explosions. Such explosions are possible if a C/O WD
accretes and retains a thick He shell. A detonation in this shell can trigger
an explosion in the C/O core of the WD (e.g., \citealt{Nomoto82,Nomoto82a}).

Now the most pressing questions are the following: which binary companion WD
(DD) or nondegenerate star (SD), and which mass explosion, $M_\mathrm{Ch}$ or
sub-$M_\mathrm{Ch}$, is dominant in the production of normal SNe\,Ia?

Observing SNe\,Ia in the hours to days after explosion provides a clear
avenue toward addressing these pressing questions \citep[e.g.,][]{Maoz14}.
Finding an SN during this early phase probes the progenitor environment and
the binary companion, which is simply not possible once the SN evolves well
into the expansion phase (they are standardizable precisely because they are
all nearly identical at maximum light). Indeed, in the landmark discovery of
SN\,2011fe, \citet{Nugent11} were able to constrain the time of explosion to
$\pm 20$\,min -- though see \citet{Piro13,Piro14} for an explanation of a
potential early ``dark phase.'' \citet{Bloom12a} would later combine the
observations in \citet{Nugent11} with an early nondetection while comparing
the limits to shock-breakout models to constrain the size of the progenitor to
be $\lesssim 0.01\,R_\odot$, providing the most direct evidence to date that
SNe\,Ia come from WDs.

Early observations also probe the nature of the binary system. In the SD
scenario, the SN ejecta will collide with the nondegenerate companion,
creating a shock that gives rise to a bright ultraviolet/optical flash in the
days after explosion \citep{Kasen10a}. To date, the search for such a
signature in large samples has typically resulted in a nondetection (e.g.,
\citealt{Hayden10,Bianco11,Ganeshalingam11}).\footnote{There are claims of
companion interaction based on short-lived optical ``bumps'' in the early
light curves of individual SNe (e.g.,
\citealt{Cao15,Marion16,Hosseinzadeh17,Dimitriadis19}). Alternative models
(e.g., \citealt{Dessart14,Piro16,Levanon17,Polin19,Magee20a}) utilizing
different physical scenarios can produce similar bumps, leading many (e.g.,
\citealt{Kromer16,Noebauer17,Miller18,Shappee18,Shappee19,Miller20a}) to
appeal to alternative explanations to ejecta-companion interaction.}
Separately, for the DD scenario, some sub-$M_\mathrm{Ch}$ DD
explosions exhibit a highly unusual color evolution in the $\sim$2--4\,d after
explosion \citep{Noebauer17,Polin19}.

Furthermore, measurements of the SN rise time constrains the properties of the
WD and the explosion. In combination with the peak bolometric luminosity, the
rise time provides a direct estimate of the ejecta mass
\citep[e.g.,][]{Arnett82,Jeffery99}. Regarding the explosions, initial work to
estimate the rise times of SNe\,Ia clearly demonstrated that early models
significantly underestimated the opacities in the SN ejecta (e.g.,
\citealt{Riess99a}). Finally, while the famous luminosity--decline
relationship for SNe\,Ia makes them standardizable \citep{Phillips93}, recent
evidence suggests that the rise, rather than the decline, of SNe\,Ia is a
better indicator of their peak luminosity \citep{Hayden19}.

To date, we have not reached a consensus on the typical rise time of SNe\,Ia.
In their seminal study, \citet{Riess99a} found that the mean rise time of SNe
Ia is $19.5 \pm 0.2$\,d, after correcting the individual SNe for the
luminosity--decline relation (we hereafter refer to these corrections as shape
corrections). Follow-up studies estimated a similar mean rise time for
high-redshift SNe\,Ia \citep{Aldering00,Conley06}.\footnote{\citet{Aldering00}
importantly point out that rise time estimates can be significantly biased if
uncertainties in the time of maximum light are ignored.} In \citet{Hayden10},
\citet{Ganeshalingam11}, and \citet{Gonzalez-Gaitan12} similar approaches were
applied to significantly larger samples of SNe, and shorter (by $\gtrsim
1$\,d) shape-corrected mean rise times were found.

As observational cadences have increased over the past decade, there has been
a surge of SNe\,Ia discovered shortly after explosion. This has led more
recent efforts to focus on measuring the rise times of populations of
individual SNe (e.g., \citealt{Firth15,Zheng17a,Papadogiannakis19}), which is
the approach adopted in this study. The utility of avoiding shape corrections
is that it allows one to search for multiple populations in the distribution
of rise times, which could point to a multitude of explosion scenarios. While
\citet{Papadogiannakis19} found no evidence for multiple populations,
\citet{Ganeshalingam11} found that high-velocity SNe\,Ia rise $\sim$1.5\,d
faster than their normal counterparts.

We are now in an era where hydrodynamic radiation transport models have become
very sophisticated (e.g.,
\citealt{Sim13,Dessart14,Kromer16,Noebauer17,Polin19,Townsley19,Gronow20,Magee20}). Accurate measurements of the observed distribution of SN Ia rise times
can be compared with models to rule out theoretical scenarios that evolve too
quickly or too slowly (e.g., \citealt{Magee18}). Similarly, if the early
emission is modeled as a power law in time (i.e., $f \propto t^\alpha$),
measures of the power-law index $\alpha$ can confirm or reject different
explosion/ejecta-mixing scenarios \citep{Magee20}.

In this paper, the second in a series of three examining the photometric
evolution of 127 SNe\,Ia with early observations discovered by the Zwicky
Transient Facility (ZTF; \citealt{Bellm19,Graham19,Dekany20}) in 2018, we
examine the rise time of SNe\,Ia and whether or not their early emission can be
characterized as a simple power law. Paper I \citep{Yao19} describes the
sample, while Paper III \citep{Bulla20} discusses the color evolution
of SNe\,Ia shortly after explosion. The sample, which is large, is equally
impressive in its quality: ZTF observations are obtained in both the \gztf\
and \rztf\ filters every night. The nightly cadence and multiple
filters are essential to constrain the distribution of $^{56}$Ni in the SN
ejecta (e.g., \citealt{Magee20}). Conducting the search with the same
telescope that provides follow-up observations enables subthreshold
detections \citep{Yao19}, separating the ZTF sample from other low-$z$ data
sets. We construct a Bayesian framework to estimate the rise time of
individual SNe. This approach allows us to naturally incorporate
priors into the model fitting. We uncover a systematic bias whereby the rise
times of the higher-redshift SNe within a flux-limited survey are typically
underestimated. We show that the adoption of strong priors can, at least
partially, alleviate this bias. Finally, we conclude with a discussion of
lessons from ZTF that can be applied to the Vera C.~Rubin Observatory
Legacy Survey of Space and Time (LSST).

Along with this paper, we have released our open-source analysis for this
study. It is available online at
\href{https://github.com/adamamiller/ztf_early_Ia_2018}
{\url{https://github.com/adamamiller/ztf_early_Ia_2018}}.

\section{ZTF Photometry}\label{sec:ztf}

\citet{Yao19} provide detailed selection criteria for the 127 SNe\,Ia utilized
in Papers I, II, and III in this series on early observations of ZTF SNe\,Ia.
In summary, these 127 SNe were observed as part of the high-cadence
extragalactic experiment conducted by the ZTF partnership in 2018
\citep{Bellm19a}. This experiment monitors $\sim$3000\,$\deg^2$ on a nightly
basis (over the nine month period when the fields are visible), with the aim
of obtaining three \gztf\ and three \rztf\ observations every night. In total,
there were 247 spectroscopically confirmed SNe\,Ia discovered within these
fields, as tabulated by the GROWTH Marshal \citep{Kasliwal19}.
Following cuts to limit the sample to SNe that were discovered ``early''
(defined as 10\,d or more, in the SN rest frame, prior to the time of maximum
light in the $B$ band, \tbmax) and have high-quality light curves, the sample
was reduced to 127 SNe; see \citep{Yao19} for the full details.

In \citet{Yao19}, we produced ``forced'' point-spread function (PSF)
photometry for each of the 127 SNe on every image covering the position of the
SN. The PSF model was generated as part of the ZTF real-time image subtraction
pipeline \citep{Masci19}, which uses an image-differencing technique based on
\citet{Zackay16}. The forced PSF photometry procedure fixes the position of
each SN and measures the PSF flux in all images that contain the SN position,
even in epochs where the signal-to-noise ratio (S/N)\,$\la 1$ and the SN is
not detected.

We normalize the SN flux relative to the observed peak flux in the \gztf\ and
\rztf\ bands as measured by \texttt{SALT2} \citep{Guy07}; see
\citet{Yao19} for our \texttt{SALT2} implementation details. The relative
fluxes produced via this procedure are unique for every ZTF reference image
(hereafter fcqf\,ID following the nomenclature in \citet{Yao19}). The ZTF
field grid includes some overlap, and SNe that occur in overlap regions will
have multiple fcqf\,IDs for a single filter. Estimates of the baseline flux
(see Equation~\ref{eqn:flux_model}) must account for the individual
fcqf\,IDs. 

The final photometric measurements in \citet{Yao19} utilize estimates of the
baseline flux, $C$, and $\chi^2_{\nu}$, which account for initially
underestimated uncertainties, to correct the results of the forced PSF
photometry. For this study, we do not employ the corrections suggested in
\citet{Yao19}. Instead, we incorporate these values directly into our model so
they can be marginalized over and effectively ignored.

\section{A Search for Early Optical Flashes}\label{sec:flash}

We constrain the presence of extreme optical flashes in SNe\,Ia by
searching our sample for sources with light curves that initially decline
before following the typical rise of an SN\,Ia. While this simple criterion
excludes events that exhibit an early bump, such as SNe\,2017cbv and 2018oh,
ZTF has found SNe\,Ia with an early optical flash (e.g., SN\,2019yvq; see
\citep{Miller20a}). Furthermore, the detection of bumps, as opposed to
``flashes'' where the flux is observed to decline, requires physical models of
the early emission \citep[e.g.,][]{Levanon17}, and cannot be captured by the
empirical power-law models used in this study. A comparison of the light
curves to explosion models, as well as an estimate of the rate of potential
flux excesses in the early light curves, will be presented in a future study
(M. Deckers et al.\ 2020, in preparation).

Of the 127 SNe in our sample, only two (ZTF18abklljv and ZTF18abptsco)
show a decline in flux following the epoch of discovery. When accounting for
the uncertainties on the individual flux measurements, however, the observed
decline in both ZTF18abklljv (SN\,2018lpk) and ZTF18abptsco (AT\,2018lpm) is
statistically consistent with a constant flux during the first two nights of
detection. We therefore conclude that an early optical flash is not detected
in any of the SNe in our sample.

Despite this lack of detection, we can still constrain the rate of such
events using binomial statistics. With no detections in 127 SNe, a naive
estimate of the rate of optical flashes in SNe\,Ia is $\lesssim$3\%. However,
not every SN in our sample is discovered sufficiently early to rule out the
presence of a flash. Using the distance moduli from \citet{Yao19} and the
host-galaxy reddening and $K$ corrections from \citet{Bulla20}, we split our
sample based on the absolute magnitude at the epoch of discovery, and use
these subsets to constrain the rate of optical flashes in SNe\,Ia as
summarized in Table~\ref{tab:flash}.

\begin{deluxetable}{lcccc}[htp]
\tablecaption{Upper Limits on the Rate of Optical Flashes in SNe\,Ia \label{tab:flash}}
\tablehead{
\colhead{}
& \colhead{}
& \colhead{}
& \multicolumn{2}{c}{Flash Fraction} \\
\cline{4-5}
\colhead{$M_\mathrm{disc}$}
& \colhead{~~~$N_\mathrm{SN}$~~~}
& \colhead{~~~$N_\mathrm{flash}$~~~}
& \colhead{~~~C\&P~~~}
& \colhead{~~~Jeffreys~~~}
} 
\startdata
$> -16.5$\,mag & 33 & 0 & $<0.11$ & $<0.07$ \\
$> -16.0$\,mag & 15 & 0 & $<0.22$ & $<0.15$ \\
$> -15.5$\,mag & 8 & 0 & $<0.37$ & $<0.26$ \\
\enddata
\tablecomments{$N_\mathrm{SN}$ is the number of SNe with an absolute magnitude
at the epoch of discovery ($M_\mathrm{disc}$) fainter than the given cuts
($-16.5$, $-16.0$, $-15.5$\,mag) in both the \gztf\ and \rztf\ bands.
$N_\mathrm{flash}$ is the number of SNe with observed flashes. The flash
fraction represents the 95\% confidence interval upper limit on the rate of
early flashes from SNe\,Ia. It has been calculated two ways: (i) using the
method of \citet{Clopper34}, and (ii) using the Jeffreys prior \citep[see][]{Cai05}.
}
\end{deluxetable}

With only eight SNe fainter than $M = -15.5$\,mag at the epoch of
discovery, we find a limit on faint optical flashes of $\lesssim$30\%. Flashes
this faint are expected when the SN ejecta collide with main-sequence
companions \citep[e.g.,][]{Kasen10a}. This limit is not that constraining
given that the companion-interaction signature is only expected in $\sim$10\%
of SNe \citep{Kasen10a}.

\section{Modeling the Early Rise of SNe\,Ia}\label{sec:model}

Following arguments first laid out in \citet{Riess99a}, the rest-frame optical
flux of an SN Ia should evolve $\propto t^2$ shortly after explosion. For an
ideal, expanding fireball, the observed flux through a passband covering the
Rayleigh-Jeans tail of the approximately blackbody emission will be $f \propto
R^2 T = v^2 t^2 T$, where $f$ is the SN flux, $T$ is the blackbody
temperature, $R$ is photospheric radius, $v$ is the ejecta velocity, and $t$
is the time since explosion.\footnote{All times reported in this study have
been corrected to the SN rest frame.} While these idealized conditions are not
perfectly met in nature, as $T$ and $v$ clearly change shortly after explosion
(e.g., \citealt{Parrent12}), their relative change is small compared to $t$.
Thus, the ``$t^2$ law'' should approximately hold, and indeed many studies of
large samples of SNe\,Ia have shown that, in the blue optical filters, $f
\propto t^2$ to within the uncertainties (e.g., \citealt{Conley06, Hayden10,
Ganeshalingam11, Gonzalez-Gaitan12}). At the same time, several recent studies
of individual, low-redshift SNe\,Ia show strong evidence that a power-law
model for the SN flux only reproduces the data if the power-law index,
$\alpha$, is significantly lower than 2 (e.g.,
\citealt{Zheng13,Zheng14,Shappee16,Miller18,Dimitriadis19,Fausnaugh19}).

For this study, we characterize the early emission in a single filter from an
SN Ia as a power law:
\begin{equation}
    f_b(t) = C + H[t_\mathrm{fl}] A_b (t - t_\mathrm{fl})^{\alpha_b},
    \label{eqn:flux_model}
\end{equation}
where $f_b(t)$ is the flux in filter $b$ as a function of time $t$ in the SN
rest frame, $C$ is a constant representing the baseline flux present in the
reference image prior to SN discovery, $t_\mathrm{fl}$ is the time of first
light,\footnote{Note that $t_\mathrm{fl}$ is not the time of explosion, but
rather the time when optical emission begins for the SN, as the observed
emission due to radioactive decay in the interior of the SN ejecta must first
diffuse through the photosphere; see e.g., \citet{Piro13,Piro14}.}
$H[t_\mathrm{fl}]$ is the Heaviside function equal to 0 for $t <
t_\mathrm{fl}$ and 1 for $t \ge t_\mathrm{fl}$, $A_b$ is a constant of
proportionality in filter $b$, and $\alpha_b$ is the power-law index
describing the rise in filter $b$.

For ZTF, observations are obtained in the \gztf\ and \rztf\ bands, and we
model the evolution in both filters simultaneously. While, strictly speaking,
$t_{\mathrm{fl}, g} \ne t_{\mathrm{fl}, r}$, we expect these values to be
nearly identical given the similarity of the SN ejecta opacity at these
wavelengths (e.g., Figure~6 in \citealt{Magee18}), and assume we cannot
resolve the difference with ZTF data. Therefore, we adopt \tfl\ ($\approx
t_{\mathrm{fl}, g} \approx t_{\mathrm{fl}, r}$) as a single parameter for our
analysis. As discussed in \citet{Yao19}, $C$ is a function of fcqf\,ID, which
represents both the filter and ZTF field ID (see also \S\ref{sec:ztf}).

Many of the SNe in our sample are observed at an extremely early phase in
their evolution, at times when the spectral diversity in SNe\,Ia is not
well-constrained; see, for example, the bottom panel of Figure~1 in
\citet{Guy07}. As a result, we do not apply $K$ corrections to the ZTF light
curves prior to model fitting. Furthermore, without precise knowledge of the
time of explosion, it is impossible to know which observations in the ZTF
nightly sequence should be corrected and which should not. While examining
SN\,2011fe, \citet{Firth15} find that ignoring $K$ corrections leads to a
systematic uncertainty on $\alpha$ of $\pm0.1$, which is smaller than the
typical uncertainty we measure (see \S\ref{sec:mean_parameters}). This
suggests that our inability to apply $K$ corrections does not significantly
affect our final conclusions.

If we assume that the observed deviations between the model flux and the data
are the result of Gaussian scatter, the log-likelihood for the data is:
\begin{equation}
    \ln \mathscr{L} \propto -\frac{1}{2}\sum_{d,i} \frac{[f_{d,i} - f_d(t_i)]^2}{(\beta_d \sigma_{d,i})^2} -\sum{\ln (\beta_d \sigma_{d,i})},
\end{equation}
where the sum is over all fcqf\,IDs $d$ and all observations $i$. Here,
$f_{d,i}$ is the $i^\mathrm{th}$ flux measurement with corresponding
uncertainty $\sigma_{d,i}$, and $\beta_d$ is a term we add to account for the
fact that the uncertainties are underestimated (see \citealt{Yao19}). Finally,
$f_d(t_i)$ is the model, Equation~\ref{eqn:flux_model} evaluated at the time
of each observation $t_i$, with $C$ replaced by $C_d$, the baseline for the
individual fcqf\,IDs, and $A_b$ and $\alpha_b$ replaced by $A_{b\mid d}$ and
$\alpha_{b\mid d}$, respectively, as these terms depend on fcqf\,ID, but only
the filter $b$ and not the field ID.

Ultimately, we only care about three model parameters: \tfl, and the power-law
index describing the rise in the \gztf\ and \rztf\ filters, hereafter
$\alpha_g$ and $\alpha_r$, respectively. Following Bayes' Law, we multiply the
likelihood by a prior and use an affine-invariant, ensemble Markov Chain Monte
Carlo (MCMC) technique \citep{Goodman10} to approximate the model posterior.

There is a strong degeneracy between $A_{b\mid d}$ and $\alpha_{b\mid d}$,
which we find can be removed with the following change of variables
$A^\prime_{b\mid d} = A_{b\mid d} 10^{\alpha_{b\mid d}}$ in
Equation~\ref{eqn:flux_model}. We adopt the Jeffreys prior \citep{Jeffreys46}
for the scale parameters $A_{b\mid d}$ and $\beta_d$, and wide flat priors for
all other model parameters, as summarized in Table~\ref{tab:priors}. The MCMC
integration is performed using \texttt{emcee} \citep{Foreman-Mackey13}. Within
the ensemble, we use 100 walkers, each of which is run until convergence or
three million steps, whichever comes first. We test for convergence by
examining the average autocorrelation length of the individual chains $\tau$
after every 20,000 steps. We consider the chains converged if
$n_\mathrm{steps} > 100 \,\tau$, where $n_\mathrm{steps}$ is the total number
of steps in each chain, and the change in $\tau$ relative to the previous
estimate has changed by $<1\%$.

\begin{deluxetable}{llc}[htp]
\tablecaption{Model Parameters $\theta$ and Their Priors \label{tab:priors}}
\tablehead{
\colhead{$\theta$}
& \colhead{Description}
&\colhead{Prior}
} 
\startdata
$C_d$ & Baseline flux per fcqf\,ID $d$ & $\mathcal{U}(-10^8,10^8)$ \\
$t_\mathrm{fl}$ & Time of first light & $\mathcal{U}(-100,0)$ \\
$A^\prime_{b\mid d}$ & Proportionality factor per filter $b$ & ${A^\prime_{b\mid d}}^{-1} 10^{-\alpha_{b\mid d}}$ \\
$\alpha_{b\mid d}$ & Rising power-law index per filter $b$ & $\mathcal{U}(0,10^8)$ \\
$\beta_{d}$ & Uncertainty scale factor per fcqf\,ID $d$ & $\beta_{d}^{-1}$ \\
\enddata
\tablecomments{The factor of $10^{-\alpha_{b\mid d}}$ in the prior for $A^\prime_{b\mid d}$ follows from the change of variables (see \ref{sec:prior}).
}
\end{deluxetable}

A key decision in modeling the early evolution of SNe\,Ia light curves is
deciding what is meant by ``early.'' While the simplistic power-law models
adopted here and elsewhere can describe the flux of SNe\,Ia shortly after
explosion, it is obvious that these models cannot explain the full evolution
of SNe\,Ia, as they never turn over and decay. Throughout the literature,
there are various definitions of early, ranging from some studies defining
early relative to the amount of time that has passed following the epoch of
discovery (e.g., \citealt{Nugent11,Zheng13,Miller18}), to others defining it
relative to the time of $B$-band maximum (e.g.,
\citealt{Riess99a,Aldering00,Conley06,Dimitriadis19}), while others define
early in terms of the fractional flux relative to maximum light (e.g.,
\citealt{Firth15,Olling15,Fausnaugh19}). Here, we adopt the latter definition
in order to be consistent with recent work using extremely high-cadence,
high-precision light curves from the space-based Kepler K2 \citep{Howell14}
and the Transiting Exoplanet Survey Satellite \citep{Ricker15} missions (e.g.,
\citealt{Olling15,Fausnaugh19}). As in \citet{Olling15}, we only include
observations up to 40\% of the peak amplitude of the SN.\footnote{We do this
separately in the \gztf\ and \rztf\ filters. In practice, we subtract a
preliminary estimate of the flux baseline derived from the median flux value
for all observations that occurred $>20$\,d (in the SN rest frame) before
\tbmax. We then divide all flux values by the peak flux determined in
\citet{Yao19}. Finally, we calculate the inverse-variance weighted mean flux
for every night of observations, and only retain those nights with
$f_\mathrm{mean} \le 0.4 f_\mathrm{max}$ for model fitting.} We find that this
particular choice, 40\% instead of 30\% or 50\%, does slightly affect the
final inference for the model parameters (see \ref{sec:systematics} for
further details).

Of the 127 SNe\,Ia in our sample, we find that the MCMC chains converge for
every SNe but one, ZTF18aaqnrum (SN\,2018bhs). Nevertheless, we retain it in
our sample as $n_\mathrm{steps} \approx 81 \,\tau$ after three million steps,
suggesting several independent samples within the chains (this SN is later
excluded from the sample; see \ref{sec:qa}).

\begin{figure*}
    \centering
    \includegraphics[width=5.2in]{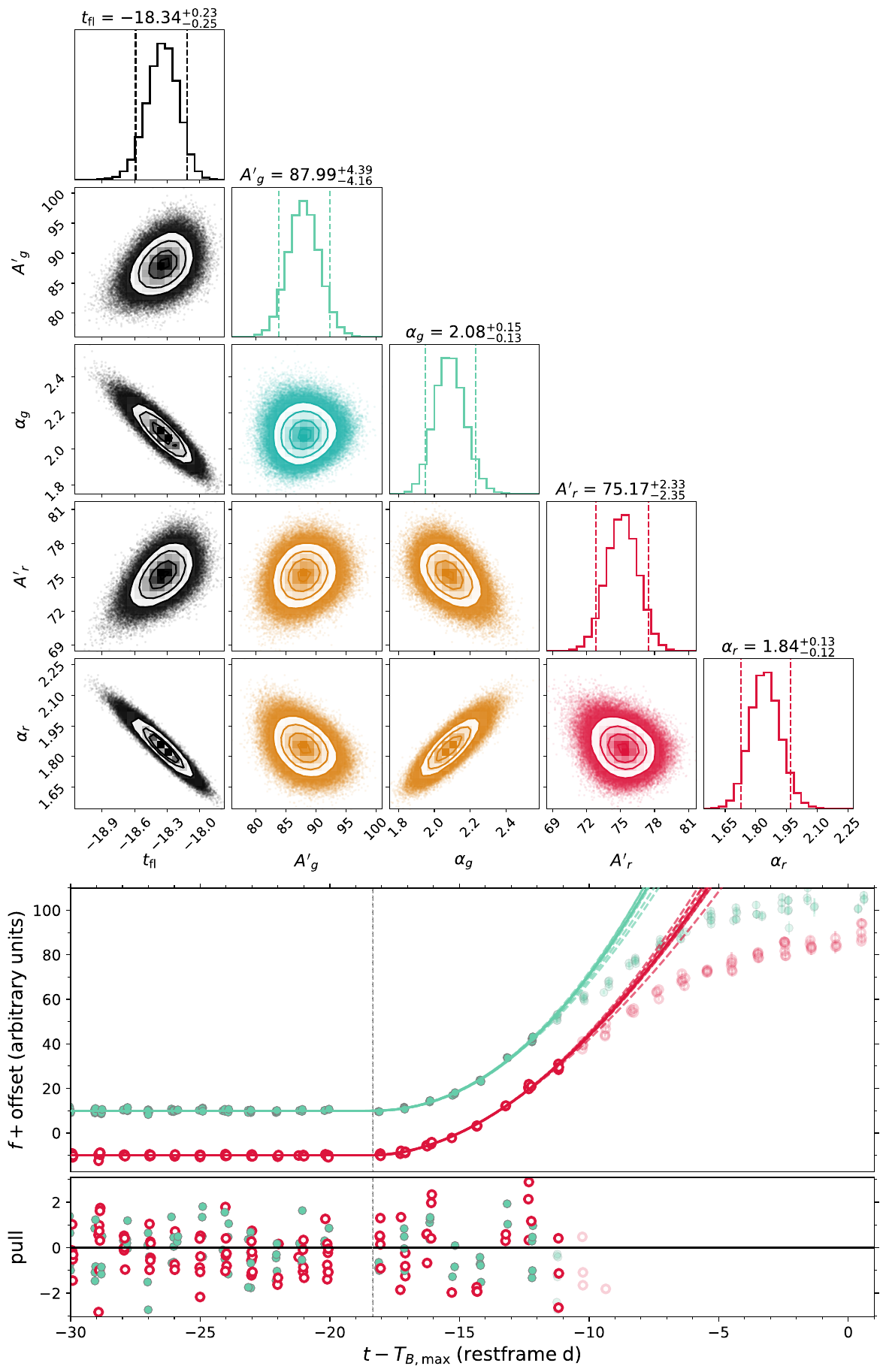}
    \caption{\textit{Top}: Corner plot showing the posterior constraints on
    \tfl, $\alpha_g$, $\alpha_r$, and the respective constants of
    proportionality, $A_g^\prime$ and $A_r^\prime$ for ZTF18abgmcmv
    (SN\,2018eay). ZTF18abgmcmv is well-fit by the model. For clarity, the
    $C_d$ and $\beta_d$ terms are excluded (in general, they do not exhibit
    strong covariance with the parameters shown here, as they are tightly
    constrained by the pre-SN observations). Marginalized one-dimensional
    distributions are shown along the diagonal, along with the median estimate
    and the 90\% credible region (shown with vertical dashed lines).
    \textit{Bottom}: ZTF light curve for ZTF18abgmcmv showing the \gztf\
    (filled, green circles) and \rztf\ (open, red circles) evolution of the SN
    in the month prior to \tbmax. Observations included in the model fitting
    (i.e., those with $f \le 0.4 f_\mathrm{max}$) are dark and solid, while
    those that are not included are faint and semi-transparent. The maximum
    \textit{a posteriori} model is shown via a thick solid line, while random
    draws from the posterior are shown with semi-transparent dashed lines. The
    vertical dashed line shows the median 1D marginalized posterior value of
    \tfl, while the thin, light gray vertical line shows \tfl\ for the maximum
    \textit{a posteriori} model (these values are nearly identical for this
    SN, and the thin gray line is not visible). The bottom panel shows the
    residuals relative to the maximum \textit{a posteriori} model normalized
    by the observational uncertainties (pull), where the factor $\beta_d$ has
    been included in the calculation of the pull.}
    \label{fig:corner_good}
\end{figure*}

\begin{figure*}
    \centering
    \includegraphics[width=5.2in]{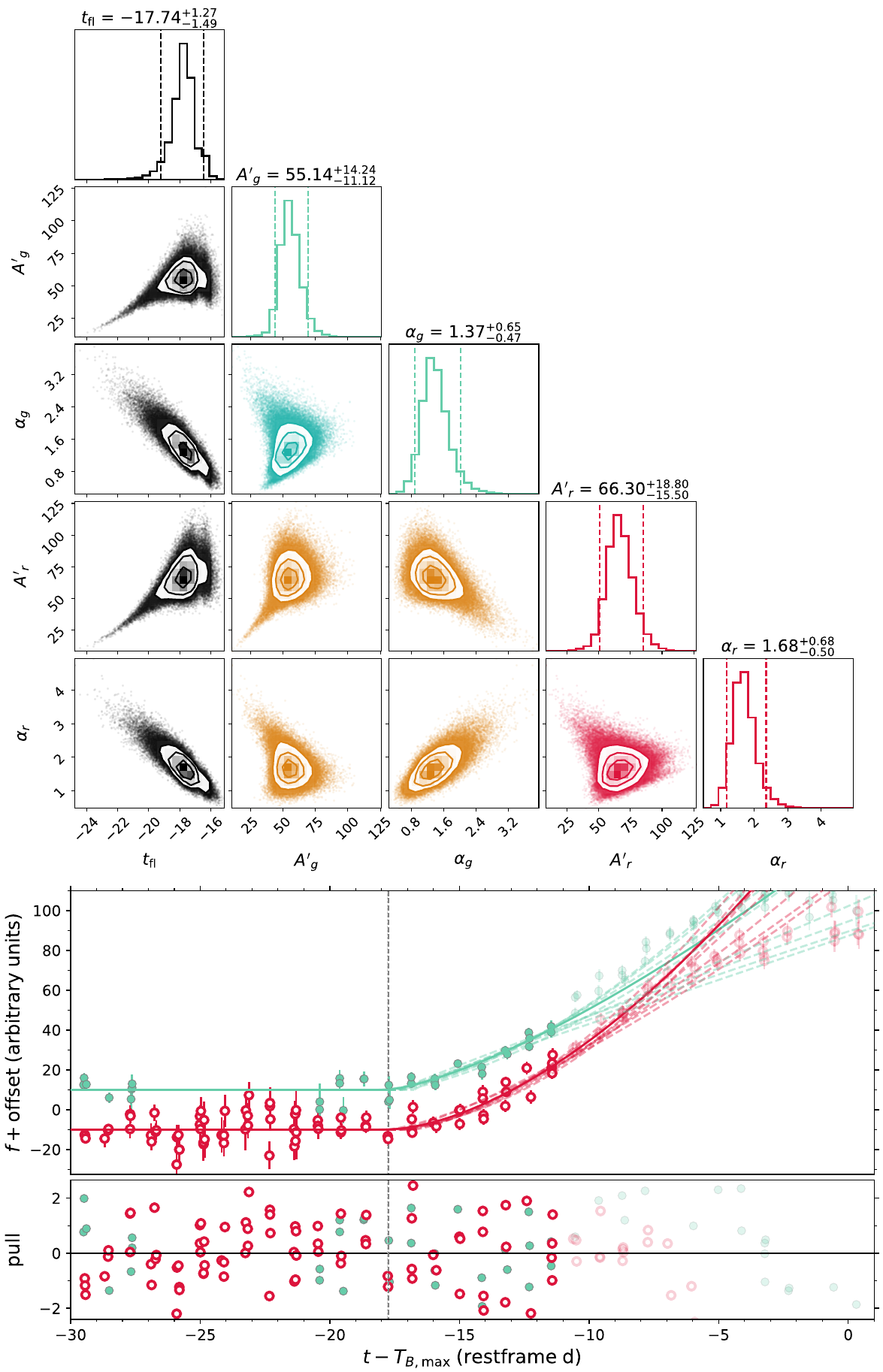}
    \caption{Same as Figure~\ref{fig:corner_good} for ZTF18abukmty
    (SN\,2018lpz), a typical SN in our sample. For ZTF18abukmty, the median 1D
    marginalized posterior value of \tfl\ and the the maximum \textit{a
    posteriori} value of \tfl\ are nearly identical, so the thin gray line
    showing the latter is not visible.}
    \label{fig:corner_median}
\end{figure*}

\begin{figure*}
    \centering
    \includegraphics[width=5.2in]{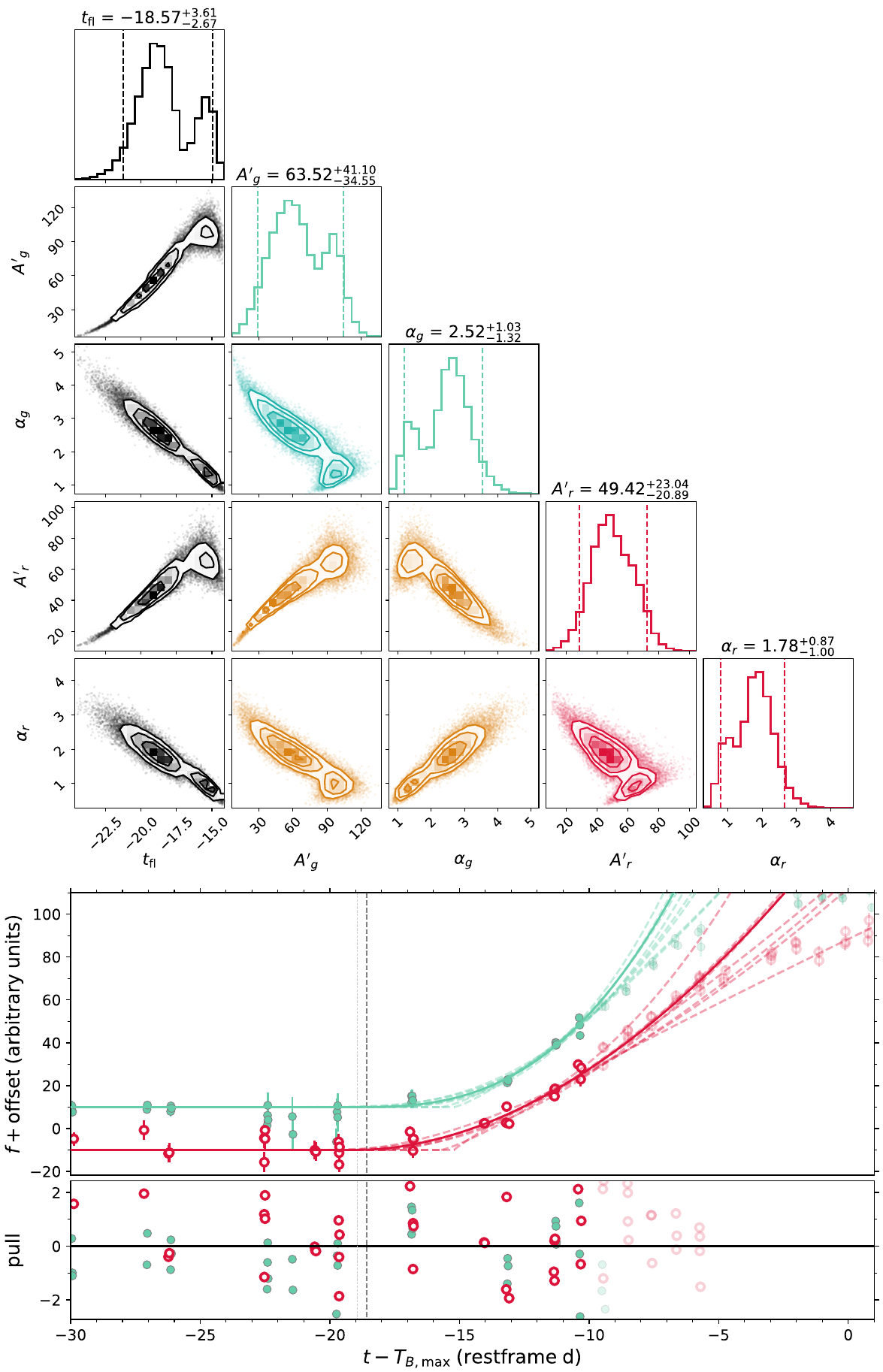}
    \caption{Same as Figure~\ref{fig:corner_good} for ZTF18aazabmh
    (SN\,2018crr), an SN that does not significantly constrain the model
    parameters.}
    \label{fig:corner_bad}
\end{figure*}

Example corner plots illustrating good, typical, and poor constraints on the
model parameters, \tfl, $\alpha_g$, and $\alpha_r$, are shown in
Figures~\ref{fig:corner_good}, \ref{fig:corner_median}, and
\ref{fig:corner_bad}, respectively. In this context, good, typical, and poor
are defined relative to the width of the 90\% credible region for \tfl\
($\mathrm{CR}_{90}$). Roughly speaking, the good models have $\mathrm{CR}_{90}
\la 1.5$\,d ($\sim$34 SNe), the median models have $\mathrm{CR}_{90} \approx
2.5$\,d ($\sim$61 SNe), and the poor models have $\mathrm{CR}_{90} \ga 4$\,d
($\sim$32 SNe). From the corner plots. it is clear that there is a positive
correlation between $\alpha_g$ and $\alpha_r$, which makes sense given the
relatively similar regions of the spectral energy distribution (SED) traced by
these filters. Finally, \tfl\ exhibits significant covariance with each of the
$\alpha$ parameters. While we report marginalized credible regions on all
model parameters in Table~\ref{tab:uninformative}, the full posterior samples
should be used for any analysis utilizing the results of our model fitting
(see, e.g., \citealt{Bulla20}).

The bottom panels of Figures~\ref{fig:corner_good}--\ref{fig:corner_bad}
display the light curves for the corresponding corner plots shown in the top
panels. In addition to the observations, we also show multiple models based on
random draws from the posterior, and the residuals normalized by the
observational uncertainties (pull) relative to the maximum \textit{a
posteriori} estimate from the MCMC sampling. As illustrated in
Figure~\ref{fig:corner_good}, we can place tight constraints on the model
parameters for light curves with a high S/N. These SNe are typically found at
low redshift, and are monitored with good sampling and at high photometric
precision. As expected, as the S/N decreases (Figure~\ref{fig:corner_median})
or the typical interval between observations increases
(Figure~\ref{fig:corner_bad}), it becomes more and more difficult to place
meaningful constraints on \tfl\ or $\alpha$. We visually examine posterior
models for each light curve and flag those that produce unreliable parameter
constraints. We use this subset of sources to identify SNe that should be
excluded from the full sample analysis described in \ref{sec:mean_parameters}
below (see \ref{sec:qa}). These flagged sources are noted in
Table~\ref{tab:uninformative}.

\section{The Mean Rise Time and Power-law Index for SNe
Ia}\label{sec:mean_parameters}

Below, we examine the results from our model fitting procedure to investigate
several photometric properties of \textit{normal} SNe\,Ia. We define normal
via the spectroscopic classifications presented in \citet{Yao19}. SNe
classified as SN\,1986G-like, SN\,2002cx-like, Ia-CSM, and super-Chandrasekhar
explosions are excluded from the analysis below. These seven peculiar events
are discussed in detail in \ref{sec:rare}. The remaining 120 normal SNe\,Ia in
our sample have $-2 \la x_1 \la 2$, where $x_1$ is the \texttt{SALT2} shape
parameter, which is well within the range of SNe that are typically used for
cosmography (e.g., \citealt{Scolnic18a}). Estimates of \trise\ and $\alpha$
for all SNe in our sample, including the seven peculiar SNe\,Ia discussed in
\ref{sec:rare}, are presented in Table~\ref{tab:uninformative}.

\subsection{Mean Rise Time of SNe\,Ia}\label{sec:mean_rise}

From the marginalized 1D posteriors for \tfl, we can examine the typical rise
time for SNe\,Ia. The model given in Equation~\ref{eqn:flux_model} constrains
\tfl, yet what we ultimately care about is the rise time, \trise. We measure
\tfl\ relative to \tbmax, which itself has some measurement
uncertainty.\footnote{In this study, \trise\ represents the rise time to
$B$-band maximum, as we measure time relative to \tbmax\ and assume \tfl\ is
the same in the $B$, \gztf, and \rztf\ filters. This assumption is
reasonable given that the opacities in an SN photosphere at frequencies
$\lesssim 10^{15}$\,Hz ($\gtrsim$ 3000\,\AA) are dominated by Thomson
scattering \citep[see Figure~6 in][]{Magee18}. Furthermore, the significant
overlap between the $B$ and \gztf\ bands suggests that \tfl\ should be
extremely similar, if not identical, in these two filters.} An estimate of
\trise\ must therefore account for the uncertainties on both \tfl\ and \tbmax.
\citet{Aldering00} critically showed that ignoring the uncertainties on the
time of maximum could lead to \trise\ estimates that are incorrect by $\ga
2$\,d.

To measure \trise, we use a Gaussian kernel density estimation (KDE) to
approximate the 1D marginalized probability density function (PDF) for \tfl.
The width of the kernel is determined via cross-validation and the KDE is
implemented with \texttt{scikit-learn} \citep{Pedregosa11}. The PDF is
multiplied by $-1$ and convolved with a Gaussian with the same variance as the
uncertainties on \tbmax\ in order to determine the final PDF for \trise. The
cumulative density function (CDF) of this PDF is used to determine the median
and 90\% credible region on \trise. We assume there is no significant
covariance in the uncertainties on \tfl\ and \tbmax. These times are estimated
using independent methods and portions of the light curve, which is why we can
convolve the uncertainties in making the final estimation of \trise.

\begin{figure}
    \centering
    \includegraphics[width=1\linewidth]{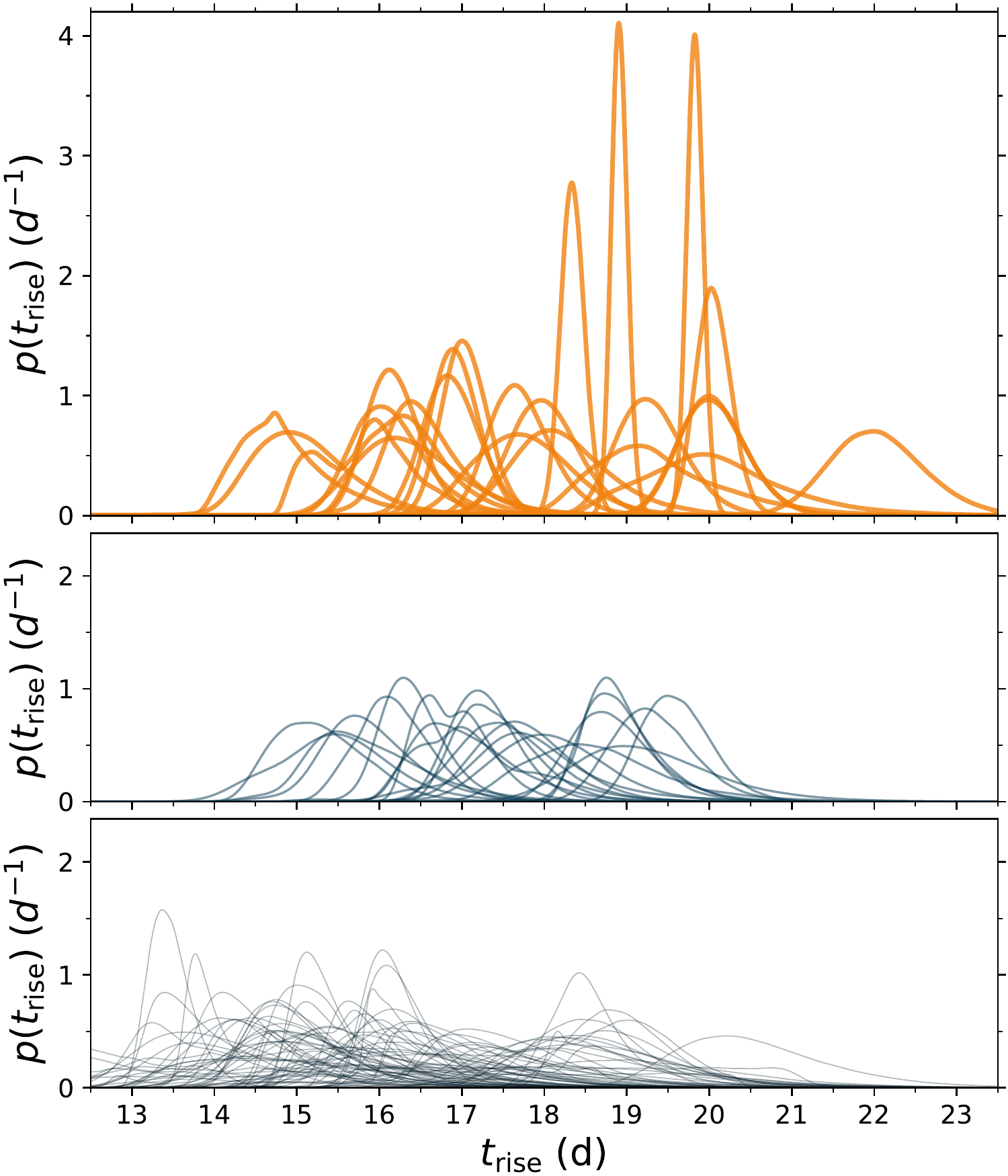}
    \caption{Marginalized posterior distribution of the rise time
    $t_\mathrm{rise}$ for normal SNe\,Ia in our sample. The sample has been
    divided into three groups (as described \ref{sec:qa}): thick
    orange lines show SNe from the reliable-$z_\mathrm{host}$ group (top
    panel), dark blue lines show SNe from the reliable-$z_\mathrm{SN}$ group
    (middle panel), and thin gray lines show SNe from the unreliable group
    (bottom panel). From the individual PDFs, it is clear that there is no
    support for a single mean $t_\mathrm{rise}$ to describe every SN in the
    sample.}
    \label{fig:rise_time}
\end{figure}

In Figure~\ref{fig:rise_time} we show the PDF for \trise\ for the 120 normal
SNe in our sample. We highlight three subsets of the normal SNe in
Figure~\ref{fig:rise_time}: SNe with reliable model parameters
(see~\ref{sec:qa}) and known host-galaxy redshifts (hereafter the
reliable-$z_\mathrm{host}$ group), SNe with reliable model parameters and
unknown host-galaxy redshifts (hereafter the reliable-$z_\mathrm{SN}$ group;
the reliable-$z_\mathrm{host}$ and reliable-$z_\mathrm{SN}$ groups together
form the reliable group), and SNe with large uncertainties in the model
parameters, typically due to sparse sampling around \tfl\ or low photometric
precision (hereafter the unreliable group).

Figure~\ref{fig:rise_time} shows that \trise\ is typically several days
shorter for SNe in the unreliable group relative to SNe in the reliable group.
This provides another indication that the low-quality light curves in our
sample are insufficient for constraining the model parameters.
Figure~\ref{fig:rise_time} also reveals that the rise time among individual
SNe does not tend toward a common mean value. If all SNe\,Ia could be
described with a single rise time, we could estimate that mean value by
multiplying the individual PDFs in Figure~\ref{fig:rise_time}
together. This product provides no support for a single rise time to describe
all SNe\,Ia (i.e., it is effectively equal to zero everywhere).

As a population, SNe\,Ia have a mean \trise\,$\approx\,17.9$\,d, where
we have estimated this value by taking a weighted mean of the median value of
the \trise\ PDFs, with weights equal to the square of the inverse of the 68\%
credible region. The mean \trise\ increases to $\approx 18.5$ and
$18.7$\,d when considering only the reliable and reliable-$z_\mathrm{host}$
subsamples, respectively (see Table~\ref{tab:mean_params}). The scatter,
estimated via the sample standard deviation, about these mean values is
$\sim$1.8\,d.

\begin{deluxetable*}{lllccccccccccccccccc}
\tabletypesize{\scriptsize}
\tablewidth{0pt}
\tablecaption{Ninety Percent Credible Regions for Marginalized Model Parameters (Uninformative Prior)\label{tab:uninformative}}
\tablehead{
\colhead{}
& \colhead{}
& \colhead{}
& \multicolumn{3}{c}{$t_\mathrm{rise}$\,(d)}
& \colhead{}
& \multicolumn{3}{c}{$\alpha_g$}
& \colhead{}
& \multicolumn{3}{c}{$\alpha_r$}
& \colhead{}
& \multicolumn{3}{c}{$\alpha_r - \alpha_g$}
& \colhead{}
& \colhead{} \\
\cline{4-6}
\cline{8-10}
\cline{12-14}
\cline{16-18}
\colhead{ZTF Name}
& \colhead{TNS Name}
& \colhead{$z$\tablenotemark{\scriptsize{a}}}
& \colhead{5}
& \colhead{50}
& \colhead{95}
& \colhead{}
& \colhead{5}
& \colhead{50}
& \colhead{95}
& \colhead{}
& \colhead{5}
& \colhead{50}
& \colhead{95}
& \colhead{}
& \colhead{5}
& \colhead{50}
& \colhead{95}
& \colhead{Reliable\tablenotemark{\scriptsize{b}}}
& \colhead{Normal\tablenotemark{\scriptsize{c}}}
}
\startdata
ZTF18aailmnv & SN\,2018ebo & 0.080 & 14.23 & 14.91 & 16.29 &  & 0.69 & 1.05 & 1.81 &  & 0.41 & 0.74 & 1.36 &  & -0.86 & -0.34 & 0.09 & n &y \\
ZTF18aansqun & SN\,2018dyp & 0.0597 & 12.45 & 13.69 & 16.33 &  & 1.15 & 3.24 & 5.66 &  & 0.23 & 0.73 & 1.88 &  & -3.06 & -2.81 & -1.46 & n &y \\
ZTF18aaoxryq & SN\,2018ert & 0.0940 & 13.30 & 14.06 & 15.42 &  & 0.31 & 0.64 & 1.10 &  & 0.14 & 0.41 & 0.84 &  & -0.65 & -0.22 & 0.20 & n &y \\
ZTF18aapqwyv & SN\,2018bhc & 0.0560 & 14.02 & 15.07 & 17.05 &  & 1.61 & 2.55 & 4.28 &  & 0.54 & 1.52 & 3.31 &  & -1.98 & -0.97 & 0.03 & n &y \\
ZTF18aapsedq & SN\,2018bgs & 0.0720 & 17.56 & 18.54 & 19.73 &  & 1.62 & 2.05 & 2.62 &  & 1.61 & 3.20 & 5.79 &  & -0.00 & 1.47 & 3.18 & n &y \\
ZTF18aaqcozd & SN\,2018bjc & 0.0732 & 10.85 & 12.13 & 16.53 &  & 0.56 & 2.28 & 4.61 &  & 1.13 & 3.47 & 5.46 &  & 0.63 & 0.81 & 0.99 & n &y \\
ZTF18aaqcqkv & SN\,2018lpc & 0.1174 & 13.16 & 14.79 & 15.99 &  & 0.62 & 2.51 & 5.21 &  & 0.21 & 1.64 & 3.69 &  & -1.74 & 0.80 & 1.82 & n &y \\
ZTF18aaqcqvr & SN\,2018bvg & 0.0716 & 13.69 & 14.32 & 15.62 &  & 0.41 & 0.69 & 1.32 &  & 0.52 & 0.89 & 1.73 &  & 0.00 & 0.25 & 0.52 & n &y \\
ZTF18aaqcugm & SN\,2018bhi & 0.0619 & 13.79 & 15.10 & 17.06 &  & 1.23 & 2.00 & 3.00 &  & 1.01 & 1.65 & 2.49 &  & -0.74 & -0.33 & 0.02 & n &y \\
ZTF18aaqffyp & SN\,2018bhr & 0.070 & 11.76 & 16.21 & 19.98 &  & 0.03 & 0.31 & 1.35 &  & 0.02 & 0.23 & 1.10 &  & -1.15 & -0.06 & 0.83 & n &y \\
ZTF18aaqnrum & SN\,2018bhs & 0.066 & 11.64 & 14.52 & 17.75 &  & 0.14 & 1.64 & 2.97 &  & 0.36 & 2.88 & 4.59 &  & -2.20 & 0.53 & 2.68 & n &y \\
ZTF18aaqqoqs & SN\,2018cbh & 0.082 & 18.31 & 18.85 & 19.67 &  & 1.08 & 1.33 & 1.72 &  & 1.06 & 1.39 & 1.88 &  & -0.17 & 0.06 & 0.33 & y &y \\
ZTF18aarldnh & SN\,2018lpd & 0.1077 & 14.07 & 15.14 & 17.17 &  & 1.20 & 2.22 & 4.21 &  & 0.77 & 1.33 & 2.37 &  & -1.71 & -1.20 & -0.37 & n &y \\
ZTF18aarqnje & SN\,2018bvd & 0.117 & 14.65 & 16.55 & 18.23 &  & 1.27 & 1.97 & 3.24 &  & 0.63 & 1.36 & 2.62 &  & -1.53 & -0.62 & 0.27 & n &y \\
ZTF18aasdted & SN\,2018big & 0.0181 & 18.76 & 18.91 & 19.08 &  & 1.46 & 1.54 & 1.63 &  & 1.30 & 1.39 & 1.50 &  & -0.21 & -0.15 & -0.09 & y &y \\
\enddata
\tablecomments{
The table includes the $5^\mathrm{th}$, $50^\mathrm{th}$, 
and $95^\mathrm{th}$ percentiles for the four parameters of interest: 
$t_\mathrm{rise}$, $\alpha_g$, $\alpha_r$, $\alpha_r - \alpha_g$. 
The 90\% credible regions are obtained by subtracting the $5^\mathrm{th}$ percentile 
from the $95^\mathrm{th}$ percentile. Estimates for $t_\mathrm{rise}$ come from 
$t_\mathrm{fl}$ while accounting for the uncertainties on $t_{B,\mathrm{max}}$, 
while estimates for the $\alpha$ parameters have been corrected to a flat prior
(see text for further details).}

\tablenotetext{a}{Redshifts are reported to four decimal places, 
if the SN host galaxy redshift ($z_\mathrm{host}$) is known. 
Otherwise, the SN redshift ($z_\mathrm{SN}$) is reported to three decimal places; 
see \citet{Yao19} for further details.}
\tablenotetext{b}{Flag for SNe with reliable model parameters (see \ref{sec:qa}):  
y = reliable group, and n = unreliable group (see text).}
\tablenotetext{c}{Flag for normal SNe\,Ia:
y = normal, and n = peculiar (the seven peculiar SNe Ia in our sample are discussed in \ref{sec:rare}; their rise times are measured relative to $T_{g,\mathrm{max}}$).}
(This table is available in its entirety in a machine-readable 
form in the online journal.)
\end{deluxetable*}

\subsection{Mean Power-law Index of the Early Rise}

We use a similar procedure to report the PDF of $\alpha_g$ and $\alpha_r$
under the assumption of a flat prior. The posterior samples for $\alpha$ shown
in Figure~\ref{fig:corner_good}, \ref{fig:corner_median}, and
\ref{fig:corner_bad} include a factor of $10^{-\alpha}$ following the
change of variables from $A$ to $A^\prime$ (see Table~\ref{tab:priors} and
\ref{sec:qa}). To remove this factor, we estimate the 1D marginalized PDF of
$\alpha$ using a KDE as above. This PDF is next divided by $10^{-\alpha}$, and
then renormalized to integrate to 1 on the interval from 0 to 10. This final
normalized PDF provides an estimate of $\alpha_g$ and $\alpha_r$ assuming a
$\mathcal{U}(0,10)$ prior.

\begin{figure}
    \centering
    \includegraphics[width=1\linewidth]{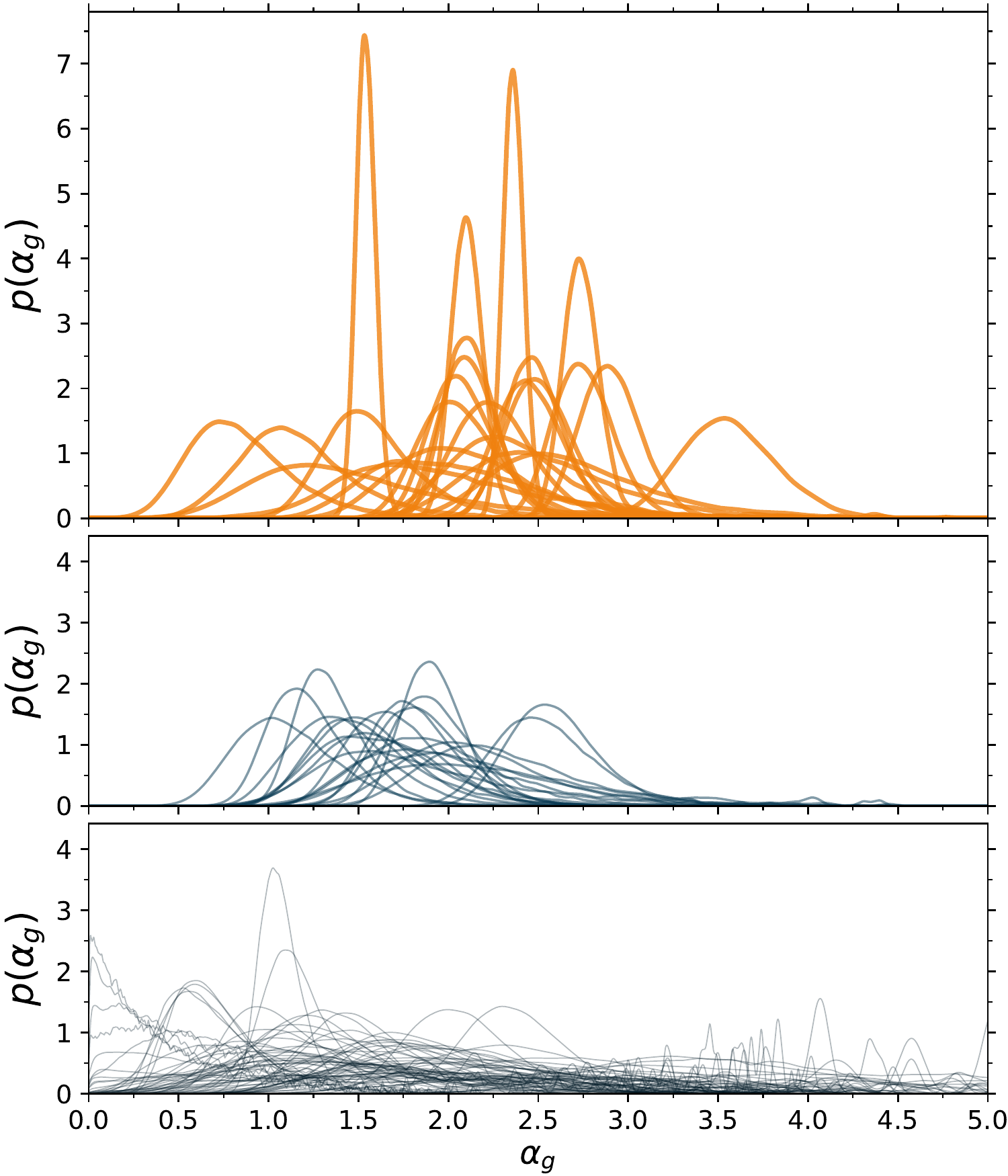}
    \caption{Marginalized posterior distribution of the rising power-law index
    in the \gztf-band, $\alpha_g$, assuming a flat prior on $\alpha_g$ for
    individual SNe in our sample. Panels and color scheme are the
    same as in Figure~\ref{fig:rise_time}. While the density of the PDFs tends
    toward 2, there is no support for a single mean power-law index to
    describe all SNe\,Ia.}
    \label{fig:alpha_rise}
\end{figure}

The PDFs for $\alpha_g$ for normal SNe\,Ia are shown in
Figure~\ref{fig:alpha_rise}. The most precise estimates of $\alpha_g$ come
from the reliable-$z_\mathrm{host}$ group, which are clustered around
$\alpha_g \approx 2$. There are, however, individual
reliable-$z_\mathrm{host}$ SNe that provide support for $\alpha_g$ as low as
$\sim$0.7 and as high as $\sim$3.5, meaning $\alpha_g$ can take on a wide
range of values.

The weighted sample mean is $\alpha_g \approx 1.9$ for normal SNe\,Ia in the
ZTF sample. This value increases to $\sim$2.1 when reducing the sample to the
reliable group or the reliable-$z_\mathrm{host}$ group. The population scatter
is $\sim$0.6 (see Table~\ref{tab:mean_params}). For $\alpha_r$ the weighted
sample mean is $\sim$1.7, $\sim$1.9, and $\sim$2.0 for the full sample, the
reliable group, and reliable-$z_\mathrm{host}$ group, respectively. The
typical scatter in $\alpha_r$ is 0.5 (see Table~\ref{tab:mean_params}). As
noted in \S\ref{sec:alpha_correlation}, there is a tight correlation
between $\alpha_g$ and $\alpha_r$, and thus we do not show the individual PDFs
for $\alpha_r$.

In both the \gztf\ and \rztf\ filters, the mean rising power-law index for the
initial evolution of the SN is close to 2, as might be expected in the
expanding fireball model. While the mean value of $\alpha$ is $\sim$2, it is
noteworthy that several SNe in the reliable-$z_\mathrm{host}$ sample are
clearly not consistent with $\alpha = 2$. If we multiply the individual PDFs
of $\alpha_g$ or $\alpha_r$ together, we find there is no support for a single
mean value of $\alpha$ for all SNe\,Ia. This suggests that models using a
fixed value of $\alpha$ are insufficient to explain the general population of
normal SNe\,Ia (though see also \S\ref{sec:strong_priors}).

\begin{deluxetable*}{lrcccccccccrcc}
\tabletypesize{\scriptsize}
\tablewidth{0pt}
\tablecaption{Population Mean and Scatter For $t_\mathrm{rise}$ and $\alpha$\label{tab:mean_params}}
\tablehead{
\colhead{}
& \colhead{}
& \multicolumn{8}{c}{Uninformative Prior}
& \colhead{}
& \multicolumn{3}{c}{$\alpha = 2$ Prior} \\
\cline{2-10}
\cline{12-14}
\colhead{}
& \colhead{}
& \colhead{$t_\mathrm{rise}$}
& \colhead{$\sigma_{t_\mathrm{rise}}$}
& \colhead{}
& \colhead{}
& \colhead{}
& \colhead{}
& \colhead{}
& \colhead{}
& \colhead{}
& \colhead{}
& \colhead{$t_\mathrm{rise}$}
& \colhead{$\sigma_{t_\mathrm{rise}}$} \\
\colhead{Subset}
& \colhead{N}
& \colhead{(d)}
& \colhead{(d)}
& \colhead{$\alpha_g$}
& \colhead{$\sigma_{\alpha_g}$}
& \colhead{$\alpha_r$}
& \colhead{$\sigma_{\alpha_r}$}
& \colhead{$\alpha_r - \alpha_g$}
& \colhead{$\sigma_{\alpha_r - \alpha_g}$}
& \colhead{}
& \colhead{N}
& \colhead{(d)}
& \colhead{(d)}
}
\startdata
Normal &                    120 & $ 17.92\pm0.04 $ & 1.81 & $ 1.89\pm0.02 $ & 0.75 & $ 1.73\pm0.02 $ & 0.80 & $ -0.18\pm0.01 $ & 0.73 && 120 & $ 21.59\pm0.02 $ & 1.89 \\
Reliable &                   47 & $ 18.53\pm0.05 $ & 1.62 & $ 2.05\pm0.02 $ & 0.53 & $ 1.89\pm0.02 $ & 0.50 & $ -0.17\pm0.01 $ & 0.23 && 115 & $ 21.59\pm0.02 $ & 1.90 \\
Reliable-$z_\mathrm{host}$ & 25 & $ 18.73\pm0.05 $ & 1.85 & $ 2.12\pm0.02 $ & 0.59 & $ 1.99\pm0.03 $ & 0.54 & $ -0.18\pm0.02 $ & 0.17 &&  58 & $ 21.54\pm0.02 $ & 2.09 \\
\hline
\multicolumn{14}{c}{Volume-limited ($z < 0.06$) subset} \\
\hline
Normal &                     28 & $ 18.61\pm0.05 $ & 2.26 & $ 2.05\pm0.03 $ & 0.76 & $ 1.95\pm0.03 $ & 0.67 & $ -0.21\pm0.01 $ & 0.86 && 28 & $ 21.68\pm0.02 $ & 1.53 \\
Reliable &                   16 & $ 18.91\pm0.05 $ & 1.75 & $ 2.13\pm0.03 $ & 0.54 & $ 2.01\pm0.03 $ & 0.52 & $ -0.18\pm0.02 $ & 0.08 && 27 & $ 21.68\pm0.02 $ & 1.51 \\
Reliable-$z_\mathrm{host}$ & 15 & $ 18.95\pm0.05 $ & 1.71 & $ 2.14\pm0.03 $ & 0.51 & $ 2.02\pm0.03 $ & 0.50 & $ -0.18\pm0.02 $ & 0.09 && 24 & $ 21.67\pm0.02 $ & 1.56 \\
DIC preferred &  & $ \ldots $ & $ \ldots $ & $ \ldots $ & $ \ldots $ & $ \ldots $ & $ \ldots $ & $ \ldots $ & $ \ldots $              && 28 & $ 19.87\pm0.03 $ & 1.46 \\
DIC--uninformative &          9 & $ 19.20\pm0.06 $ & 1.38 & $ 2.14\pm0.03 $ & 0.55 & $ 2.02\pm0.03 $ & 0.52 & $ -0.19\pm0.02 $ & 0.08 && $ \ldots $ & $ \ldots $ & $ \ldots $ \\
\enddata
\tablecomments{
Table includes the weighted mean (see text), plus standard uncertainty in the weighted mean, 
as well as the scatter (the sample standard deviation), for the four parameters of interest, 
$t_\mathrm{rise}$, $\alpha_g$, $\alpha_r$, $\alpha_r - \alpha_g$, for the uninformative 
and $\alpha = 2$ priors. $N$ is the number of SNe in each subset of the data, which are defined as follows 
(see text for more detailed definitions):
normal -- normal SNe\,Ia; 
reliable -- SNe with reliable model parameters; 
reliable-$z_\mathrm{host}$ -- reliable SNe with known host galaxy redshifts; 
DIC preferred -- results from the $\alpha = 2$ prior, 
\textit{unless the DIC prefers the uninformative prior} (see \S\ref{sec:dic}; only applies to $t_\mathrm{rise}$);
DIC--uninformative -- only SNe where the DIC prefers the uninformative prior 
(see \S\ref{sec:dic}; excludes the $\alpha = 2$ prior by construction).
The volume-limited subset includes only SNe with $z < 0.06$ (see \S\ref{sec:volume_limited}).
Note that the  definition of reliable differs for the uninformative and $\alpha = 2$ priors; 
see \ref{sec:qa} and \S\ref{sec:strong_priors}, respectively.
}

\end{deluxetable*}

\subsection{Mean Color Evolution}\label{sec:colors}

Here, we examine the mean initial color evolution of SNe\,Ia, under the
assumption that the early emission from SNe\,Ia can correctly be described by
the power-law model adopted in \S\ref{sec:model}. This analysis does not
address the initial colors of SNe\,Ia; for a more detailed analysis of the
initial colors and color evolution of SNe\,Ia, see Paper III in this series
\citep{Bulla20}.

Unlike \trise\ and $\alpha$, we do find evidence for a single mean value of
the early color evolution of SNe\,Ia, as traced by $\alpha_r - \alpha_g$. If
the early evolution in the \gztf\ and \rztf\ filters is a power law in time,
then the \gztf\ $-$ \rztf\ color, in mag, will be proportional to $(\alpha_r -
\alpha_g) \log_{10} (t - t_\mathrm{fl})$.

\begin{figure}
    \centering
    \includegraphics[width=1\linewidth]{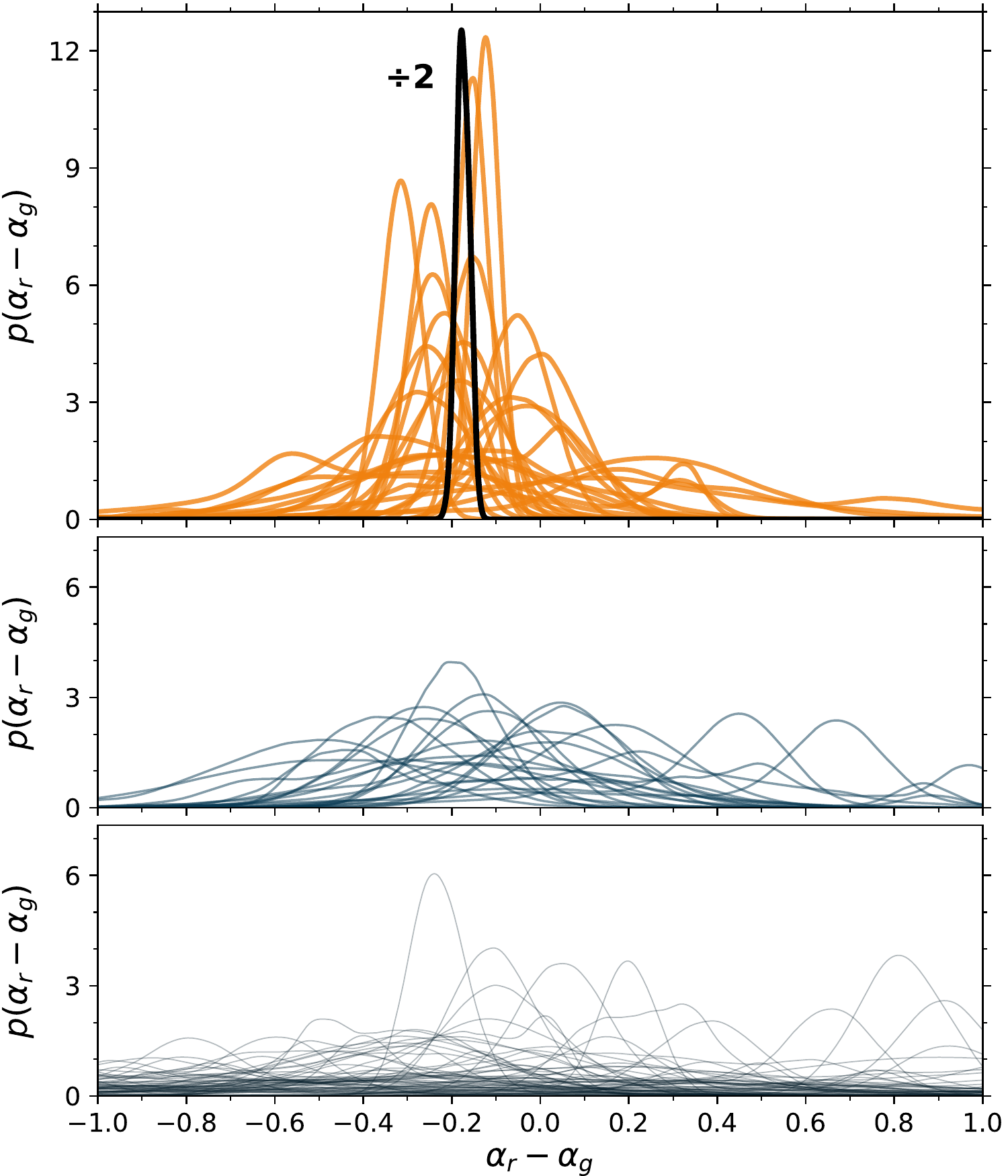}
    \caption{Marginalized posterior distribution of the early SN Ia color
    evolution, $\alpha_r - \alpha_g$, assuming flat priors on $\alpha_g$ and
    $\alpha_r$. Panels and color scheme are the same as in
    Figure~\ref{fig:rise_time}. The thick, solid black line shows an estimate
    of the mean value of $\alpha_r - \alpha_g$, which is estimated by
    multiplying together the likelihoods for SNe in the
    reliable-$z_\mathrm{host}$ group, which is why this mean is only shown in
    the top panel. For clarity, the mean PDF has been divided by 2. There is
    support for a single mean value of $\alpha_r - \alpha_g \approx -0.18$
    (see Table~\ref{tab:mean_params}).}
    \label{fig:delta}
\end{figure}

To estimate $\alpha_r - \alpha_g$ we use a similar procedure as above;
however, we need to estimate the marginalized joint posterior on $\alpha_g$
and $\alpha_r$, $\pi(\alpha_g,\alpha_r \mid t_\mathrm{fl}, A^\prime_b,
\beta_d)$, in order to correct the posterior estimates for the priors on
$\alpha$. We estimate the 2D joint posterior via a Gaussian KDE, correct this
distribution for the priors on $\alpha_g$ and $\alpha_r$, and then obtain
random draws from this distribution to estimate the 1D marginalized
likelihood estimates on $\alpha_r - \alpha_g$. The PDFs for $\alpha_r -
\alpha_g$ for individual SNe are shown in Figure~\ref{fig:delta}.

Unlike the estimates for \trise\ and $\alpha$ alone, $\alpha_r - \alpha_g$ is
clearly clustered around $\sim$${-0.2}$ for the reliable group. Multiplying
the likelihoods of the reliable-$z_\mathrm{host}$ group together produces
support for a single mean value of $\alpha_r - \alpha_g =
-0.175^{+0.016}_{-0.015}$, where the uncertainties on that estimate represent
the 90\% credible region. The mean PDF for the reliable-$z_\mathrm{host}$
group is shown as the thick, solid black line in Figure~\ref{fig:delta}. A
mean value of $\alpha_r - \alpha_g$ suggests that a typical, normal SN Ia
becomes bluer in the days after explosion. Such an evolution makes sense for
an optically thick, radioactively heated, expanding ejecta (e.g.,
\citealt{Piro16,Magee20}). There are, however, clear examples of individual
SNe that do not exhibit this behavior, e.g., SN\,2017cbv
\citep{Hosseinzadeh17} and iPTF\,16abc \citep{Miller18}, meaning this mean
behavior is not prescriptive for every SN Ia. These results exclude SNe from
the unreliable group, and their inclusion would remove any support for a
single mean value of $\alpha_r - \alpha_g$. This is largely due to a small
handful of events that feature extreme values of $\alpha_r - \alpha_g$ because
there are gaps in the observational coverage of one of the two filters (see
the upper right panel of Figure~\ref{fig:model_parameters}).

\section{Population correlations}

In addition to looking at the typical values of \trise\ and $\alpha$ for SNe
Ia, we also examine the correlations between these parameters, as well as how
they evolve with redshift, $z$. These correlations may reveal details about
the explosion physics of SNe\,Ia; for example, if strong mixing in the SN
ejecta affects the early evolution, as found in \citet{Piro16},
\citet{Magee18}, and \citet{Magee20}, then any correlations with $\alpha$ may
be related to ejecta mixing. If the model parameters are correlated with
redshift, that could be evidence for either the cosmic evolution of SNe\,Ia
progenitors or inadequacies in the model.

\begin{figure*}
    \centering
    \includegraphics[width=6in]{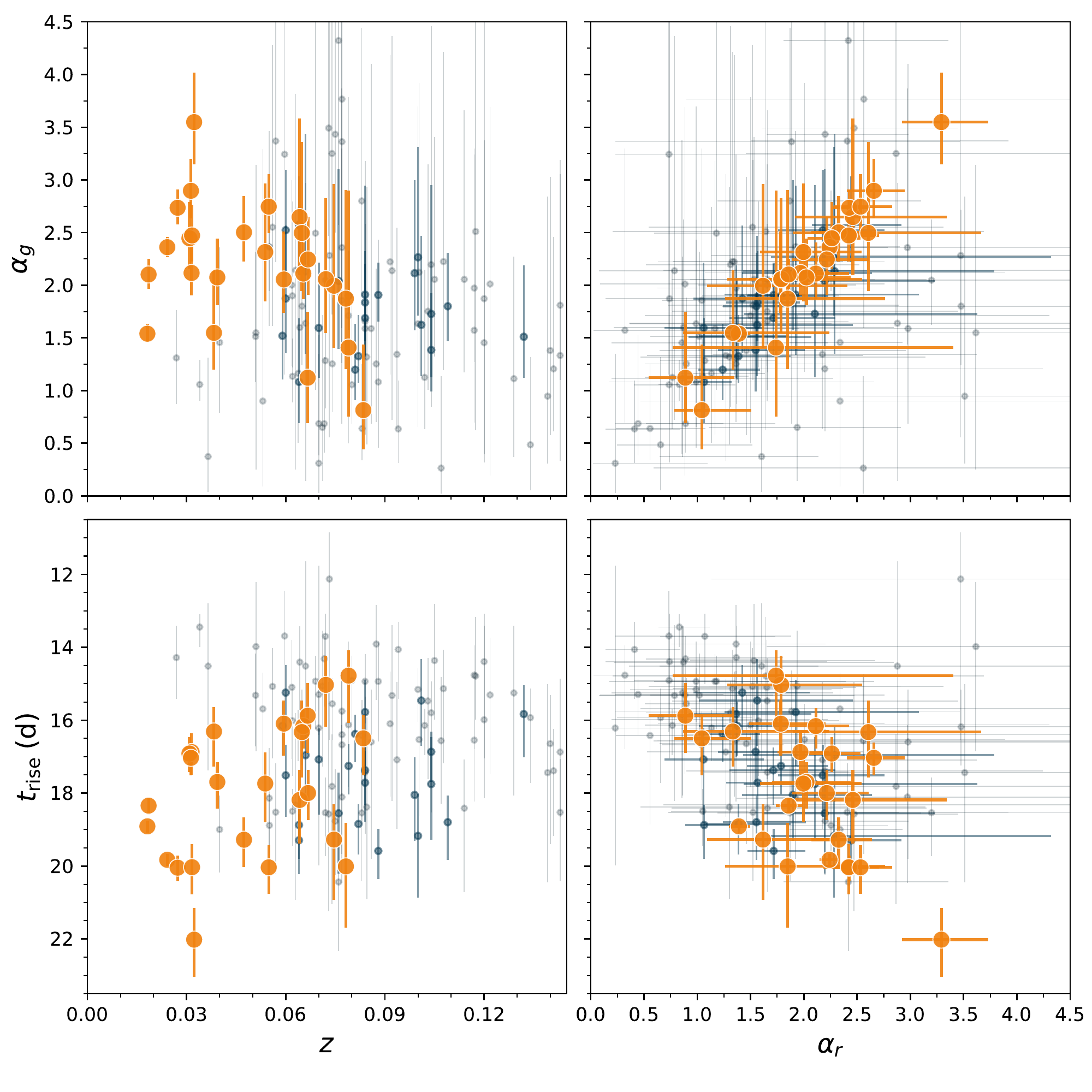}
    \caption{Correlation between redshift, $z$, SN rise time,
    $t_\mathrm{rise}$, and the power-law index in the \gztf\ and \rztf\
    filters, $\alpha_g$ and $\alpha_r$, respectively. We do not show
    $\alpha_r$ vs.~$z$ or $t_\mathrm{rise}$ vs.~$\alpha_g$, as these would
    largely be redundant given the very strong correlation between $\alpha_g$
    and $\alpha_r$ (upper right panel). The sample has been divided into three
    groups: small, light gray circles show SNe from the unreliable group (see
    \ref{sec:qa}); dark blue circles show the reliable-$z_\mathrm{SN}$ group;
    and large orange circles show the reliable-$z_\mathrm{host}$ group. The
    plots show that redshift is correlated with both $t_\mathrm{rise}$ and
    $\alpha_g$, which would only be expected if SNe\,Ia undergo significant
    evolution from $z \approx 0$ to $0.1$. We show this to be the result
    of a systematic selection effect (see text for further details). }
    \label{fig:model_parameters}
\end{figure*}

The correlation between $t_\mathrm{rise}$, $\alpha_g$, $\alpha_r$, and $z$ is
shown in Figure~\ref{fig:model_parameters}. We do not show the correlation
between $\alpha_r$ and $z$ or between $\alpha_g$ and \trise, as this
information is effectively redundant given the tight correlation between
$\alpha_g$ and $\alpha_r$ (top right panel of
Figure~\ref{fig:model_parameters}).

\subsection{Correlation between $\alpha_g$ and $\alpha_r$}\label{sec:alpha_correlation}

The most striking feature in Figure~\ref{fig:model_parameters} is the tight
correlation between $\alpha_g$ and $\alpha_r$. This result is
reasonable because the SN SED is approximately a
blackbody, and the \gztf\ and \rztf\ filters are relatively line-free
(compared to the UV) and sample adjacent portions of the Rayleigh-Jeans tail.
Thus, the evolution should be nearly identical in the two filters. SNe with
reliable model parameters follow a tight locus around $\alpha_r - \alpha_g
\approx -0.2$, with the only major outliers from this relation being SNe in
the unreliable group.

The Spearman rank-ordered correlation coefficient for $\alpha_g$ and
$\alpha_r$ is highly significant for the entire population ($\rho > 0.5$).
Restricting the sample to SNe with reliable model parameters increases the
significance of the correlation dramatically ($\rho > 0.9$). Thus, knowledge
of the power-law index in either filter provides a strong predictor for the
power-law index in the other filter.

\subsection{Correlations with Redshift -- Systematics, Not Cosmic Evolution}\label{sec:redshift_correlations}

While less prominent, Figure~\ref{fig:model_parameters} additionally shows
that both \trise\ and $\alpha$ are correlated with redshift. This result is
somewhat surprising: naively, it suggests some form of cosmic evolution in SNe
Ia, with SNe at $z \approx 0.08$ having rise times that are several days
shorter than SNe at $z \approx 0.02$. The small range of redshifts in our
sample, and several previous studies (e.g.,
\citealt{Aldering00,Conley06,Gonzalez-Gaitan12,Jones19}), render this naive
explanation in doubt. Instead, these correlations are the result of building a
sample from a flux-limited survey.

Given that ZTF cannot detect SNe when their observed brightness is
$g_\mathrm{ZTF} \ga 21.5$\,mag \citep{Masci19,Bellm19}, SNe at progressively
higher redshifts are discovered at a later phase in their evolution. The large
degeneracies in the model presented in Equation~\ref{eqn:flux_model}, namely
between \tfl, $A$, and $\alpha$, allow for a great deal of flexibility when
fitting the data. For SNe discovered at later phases, it is possible to adjust
\tfl\ while decreasing $A$ and $\alpha$, such that \tfl\ occurs around the
epoch of first detection (resulting in a shorter rise time).

\begin{figure}
    \centering
    \includegraphics[width=1\linewidth]{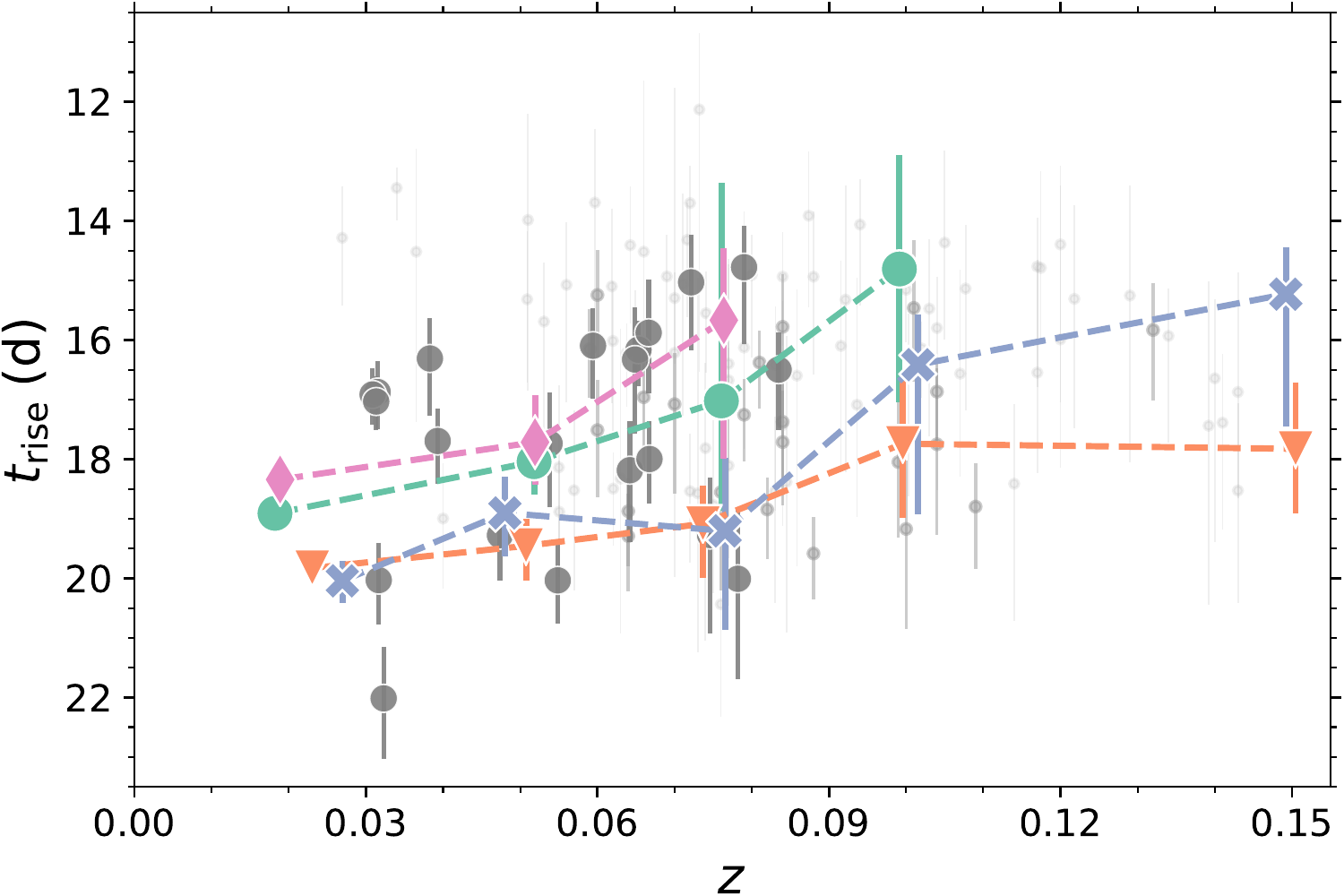}
    \caption{Same as the bottom left panel of
    Figure~\ref{fig:model_parameters}, though all SNe are shown in gray. The
    large green circle, magenta diamond, orange triangle, and purple $X$ show
    how marginalized posterior estimates of \trise\ change as the four lowest
    redshift SNe are observed at $z = 0.05$, 0.075, 0.1, and 0.15 (see text
    for further details). For clarity, slight offsets in $z$ have been applied
    to the symbols, as the error bars would otherwise fully overlap. The value
    of \trise\ clearly decreases with increasing redshift, showing that the
    observed correlation between these parameters is a consequence of
    flux-limited SN surveys.}
    \label{fig:high_z_systematic}
\end{figure}

We illustrate this effect in Figure~\ref{fig:high_z_systematic}, which shows
that the inferred rise time for identical SNe decreases as those SNe are
observed at successively higher redshifts. We use the 4 normal SNe with $z \le
0.03$ and simulate their appearance at higher redshift by making the
(over-simplified) assumption that all detections are in the sky-background
dominated regime. Thus, in any given epoch the $\mathrm{S/N} \propto
d_L^{-2}$, where $d_L$ is the SN luminosity distance. To simulate the SN at
some new redshift, $z_\mathrm{sim}$, we multiply the uncertainties by
$(d_{L,\mathrm{sim}}/d_{L,\mathrm{obs}})^2$, where $d_{L,\mathrm{sim}}$ is the
luminosity distance at $z_\mathrm{sim}$, and $d_{L,\mathrm{obs}}$ is the
observed luminosity distance to the SN.\footnote{Following \citet{Yao19}, we
adopt a flat $\Lambda$CDM cosmology with $H_0 =
73.24$\,km\,s$^{-1}$\,Mpc$^{-1}$ \citep{Riess16} and $\Omega_m = 0.275$
\citep{Amanullah10} to calculate $d_L$ for the SNe.} Using these increased
uncertainties, we randomly resample the observed flux values from a normal
distribution with mean equal to the original flux and variance equal to the
square of the distance-scaled uncertainty. After correcting the observation
times to the simulated rest frame, we fit the noisier simulated data with the
procedure from \S\ref{sec:model}. We simulate the appearance of these SNe at
redshifts $z = 0.05$, 0.075, 0.1, and 0.15. Only the models that converge are
shown in Figure~\ref{fig:high_z_systematic}.

The results shown in Figure~\ref{fig:high_z_systematic} are clear: SNe
discovered at higher redshifts have systematically smaller estimates for
\trise. This result is simple to understand as higher redshift SNe will not be
detected until later in their evolution. A stronger prior on any of the model
parameters would help to combat this effect (see \S\ref{sec:strong_priors}),
though as previously discussed we avoid strong priors due to the wide range of
$\alpha$ and \tfl\ that has been reported in the literature. 

This effect also explains the correlation seen in the bottom right panel of
Figure~\ref{fig:model_parameters}. SNe detected later in their evolution will
be evolving less rapidly as the rate of change in brightness continually
decreases until the time of maximum light. Hence, a later detection provides a
lower value of $\alpha$. Indeed, a re-creation of
Figure~\ref{fig:high_z_systematic} showing $\alpha_g$ instead of \trise\ shows
$\alpha_g$ decreasing with increasing redshift. Thus, the observed
correlations with redshift seen in Figure~\ref{fig:model_parameters} can be
entirely understood as the result of ZTF being a flux-limited survey.

The implications of this result have consequences well beyond the ZTF sample
discussed here. Essentially all SN surveys are flux-limited, meaning the
systematics associated with redshift will affect any efforts to determine
\trise\ or $\alpha$ in those data as well. The inclusion of higher-redshift
SNe in the sample will, on average, bias estimates of \trise\ and $\alpha$ to
lower values. Even more concerning is the possibility that this trend may
continue to very low redshifts ($z \ll 0.01$). The paucity of SNe in this
redshift range, due to the relatively small volume probed, make it difficult
to test for such an effect. Due to the systematic identified here, it may be
the case that the rise time, and by extension also $\alpha$, are
underestimated for every SN in the literature. Detailed simulations with
realistic SN light curves are needed to test this possibility.

\subsection{Correlation Between \trise\ and $\alpha$}

The lower right panel of Figure~\ref{fig:model_parameters} shows that \trise\
and $\alpha_r$ are correlated (and by extension \trise\ and $\alpha_g$ are
also correlated). The Spearman correlation coefficient for \trise\ and
$\alpha_r$ is significant ($\rho \gtrsim 0.5$) for both the entire population
of SNe in this study and the reliable subset as well. Similar values are found
for \trise\ and $\alpha_g$. The positive correlation between \trise\ and
$\alpha$ is also found in \citet{Gonzalez-Gaitan12}.

The origins of such a correlation may be a consequence of the $^{56}$Ni
distribution in the SN ejecta. In \citet{Magee20} a suite of models is
developed to explore the effects of $^{56}$Ni mixing on the resulting emission
from SNe\,Ia. The full family of models in \citet{Magee20}, which was designed
to cover a wide range of parameter space and not the physical space occupied
by observed SNe\,Ia, does not show significant correlation between \trise\ and
$\alpha_r$. When including only models that do a good job of reproducing
observations (see Section 5 in \citealt{Magee20}), there is a strong
correlation between \trise\ and $\alpha_r$ (Spearman $\rho > 0.9$) that
roughly matches the slope in Figure~\ref{fig:model_parameters} (M.~Magee 2020,
private communication). It is therefore possible that the observed correlation
between \trise\ and $\alpha_r$ reflects the distribution of $^{56}$Ni produced
by thermonuclear explosions.

\subsection{Correlations with Light-curve Shape}

\begin{figure*}
    \centering
    \includegraphics[width=6in]{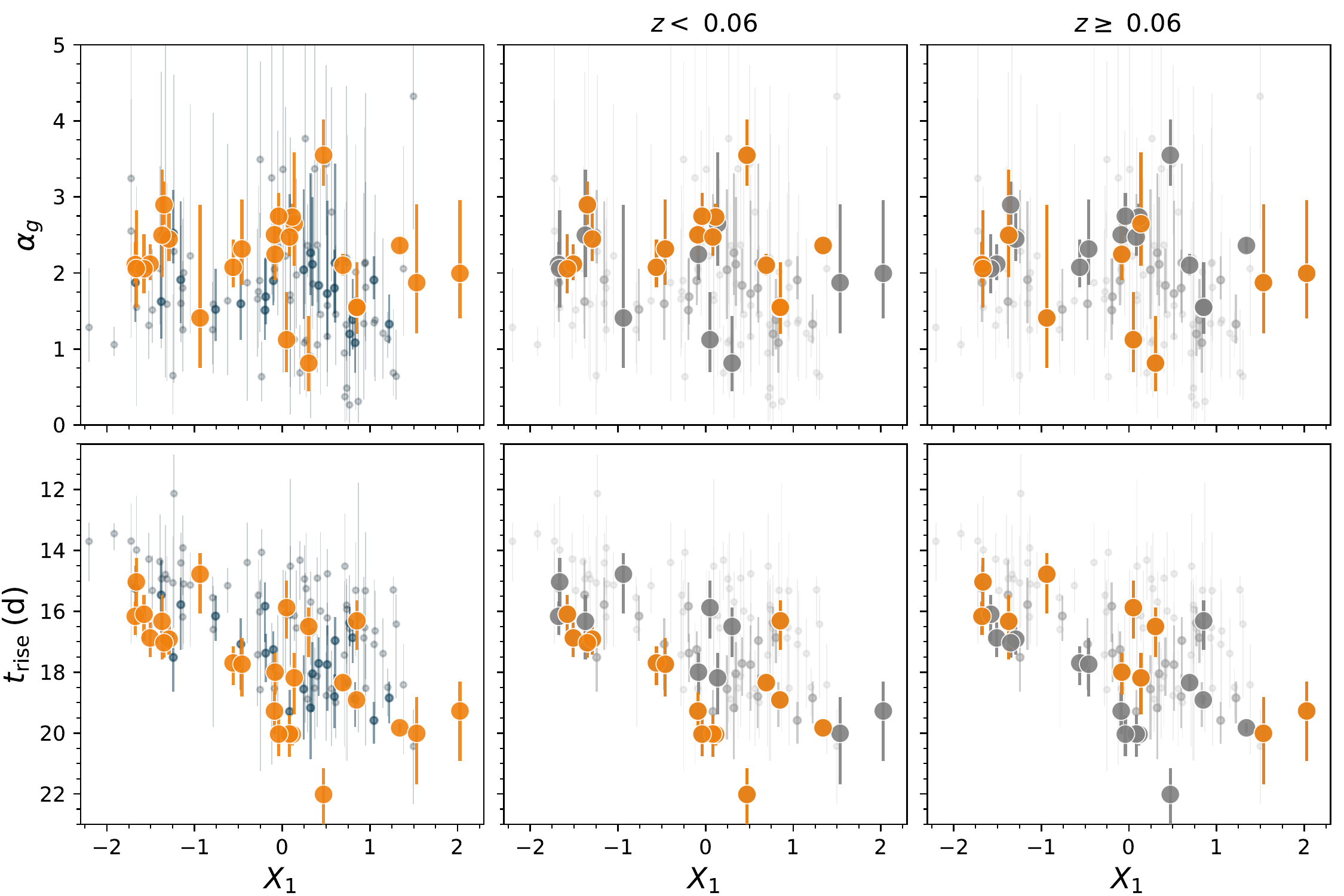}
    \caption{Correlation between the \texttt{SALT2} $x_1$ shape parameter and
    $\alpha_g$ (top row) and \trise\ (bottom row). Symbols are the same as in
    Figure~\ref{fig:model_parameters}. For clarity, the uncertainties on $x_1$
    are not shown. Here, \trise\ shows a strong correlation with $x_1$, while
    there is no correlation between $\alpha_g$ and $x_1$. The middle and right
    panels highlight reliable-$z_\mathrm{host}$ SNe at low, $z < 0.06$, and
    high, $z \ge 0.06$, redshift, respectively. Dividing the sample into
    different redshift bins shows that some of the observed scatter between
    $x_1$ and the model parameters is due to redshift and not intrinsic
    scatter. }
    \label{fig:shape_correlations}
\end{figure*}

A defining characteristic of SNe\,Ia is that they can be described by a
relatively simple luminosity--shape relation \citep{Phillips93}, which enables
them to be used as standardizable candles. We examine the correlation between
light-curve shape, in this case the \texttt{SALT2} $x_1$ parameter, and the SN
rise time and $\alpha$ in Figure~\ref{fig:shape_correlations}. There is a
clear correlation between shape and \trise, which has been hinted at in other
samples (e.g., \citealt{Riess99a,Gonzalez-Gaitan12,Firth15,Zheng17a}). The
Spearman coefficient for $x_1$ and \trise\ is significant for the entire
population ($\rho > 0.5$), and it increases when considering the reliable
group ($\rho > 0.6$).

The $x_1$ shape parameter accounts for the width of both the SN rise and
decline--and therefore, by definition, it should be correlated with the rise
time. The middle and right panels of Figure~\ref{fig:shape_correlations}
divide the reliable-$z_\mathrm{host}$ group into low ($z < 0.06$) and high ($z
\ge 0.06$) redshift bins. From these panels, it is clear that some of the
scatter in the $x_1$--\trise\ plane is the result of the redshift bias
discussed in \S\ref{sec:redshift_correlations}, as higher-redshift SNe have
shorter rise times at fixed $x_1$. A correction for this redshift effect would
reduce the overall scatter seen in the lower panels of
Figure~\ref{fig:shape_correlations} (see \S\ref{sec:strong_priors}).

\begin{figure*}
    \centering
    \includegraphics[width=6in]{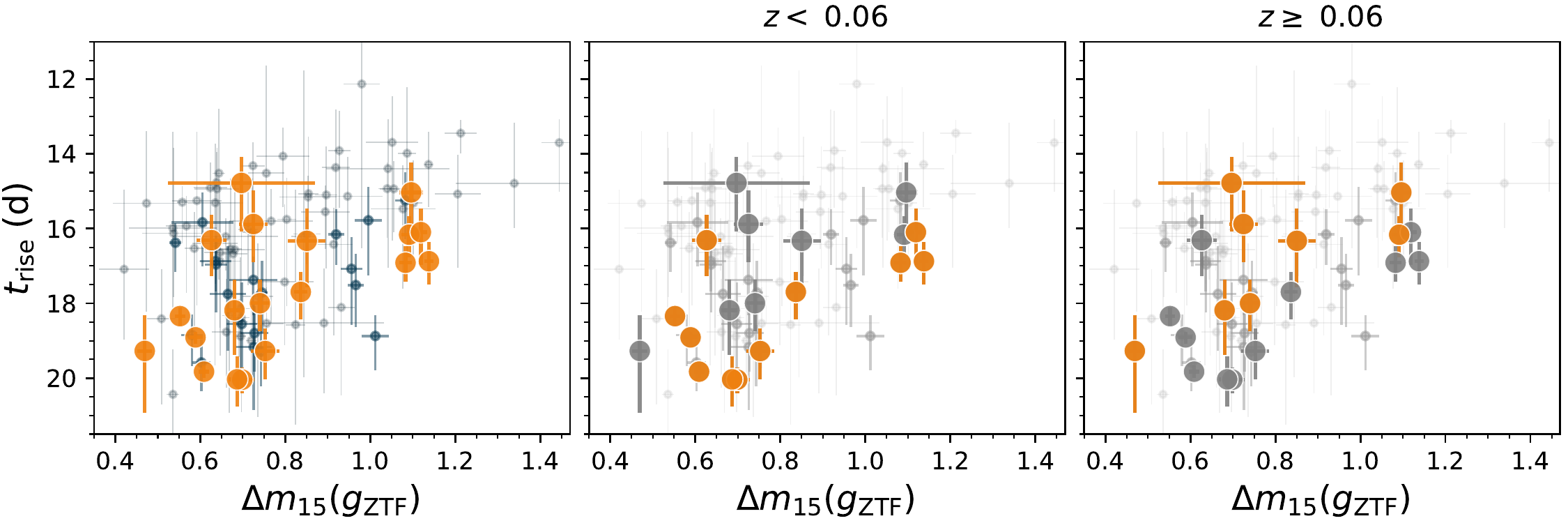}
    \caption{Correlation between $\Delta m_{15}(g_\mathrm{ZTF})$ and \trise.
    Symbols are the same as in Figure~\ref{fig:model_parameters}. There is a
    clear correlation between the rise and decline times of SNe\,Ia. Some of
    the scatter in the relationship between these parameters can be explained
    as a result of the redshift effect discussed in
    \S\ref{sec:redshift_correlations}.}
    \label{fig:dm15}
\end{figure*}

We compare separate measurements of the rise and decline of SNe\,Ia in
Figure~\ref{fig:dm15}, which shows the correlation between \trise\ and $\Delta
m_{15}(g_\mathrm{ZTF})$, the observed decline in magnitudes of the \gztf\
light curve between the time of \gztf\ maximum light and 15\,d later. The
value of $\Delta m_{15}(g_\mathrm{ZTF})$ is measured via low-order polynomial
fits to the rest-frame \gztf\ photometry ($K$ corrections have been applied
using the procedure in \citep{Bulla20}), from an SN rest-frame phase $= -7$ to
$+21$\,d. This measurement is not possible for 16 of the 127 SNe in our
sample, due to an insufficient number of observations in the defined window.
Figure~\ref{fig:dm15} shows that, on average, slowly declining SNe with
smaller values of $\Delta m_{15}(g_\mathrm{ZTF})$ have longer rise times
(Spearman $\rho \approx -0.4$ for the full sample, and $\sim$$-0.5$ for the
reliable-$z_\mathrm{host}$ group).

While \trise\ and $\Delta m_{15}(g_\mathrm{ZTF})$ exhibit a moderate
correlation, a few SNe emerge as outliers relative to this relation. Among the
low-$z$ SNe in our sample, ZTF18abkhcrj (SN\,2018emi) has $\Delta
m_{15}(g_\mathrm{ZTF}) \approx 0.6$, making it a slow-declining SN, yet the
rise time is a relatively short $\sim$16.3\,d. In this particular case, the
short rise time may be the result of larger photometric uncertainties than are
typical for ZTF SNe. ZTF18abkhcrj (SN\,2018emi) was discovered on top of the
nucleus of its host galaxy during full moon. These two effects would
contribute additional noise, delaying the phase at which the SN is
discovered--and, similar to the effects discussed in
\S\ref{sec:redshift_correlations}, yield a shorter estimate for \trise.

\citet{Hayden10} find many SNe similar to ZTF18abkhcrj, in the sense that
several of the slowest-declining SNe are among the fastest risers.
\citet{Hayden10} also find that the fastest-declining SN in their sample has
\trise\ $> 20$\,d. \citet{Gonzalez-Gaitan12} also find marginal evidence that
slow-declining SNe have faster rise times than their fast-declining
counterparts. The rise times in both \citet{Hayden10} and
\citet{Gonzalez-Gaitan12} are measured after shape correction, making it
difficult to directly compare to our ZTF sample. Nevertheless, if there is no
clear correlation, then that would suggest that the rise and fall of SNe\,Ia
is set by different physics. Additional samples should be collected to test
for this possibility.

There is no strong correlation between $\alpha$ and $x_1$
(Figure~\ref{fig:shape_correlations}). The Spearman correlation for these two
parameters is $\rho \approx -0.2$ whether looking at $\alpha_g$ or $\alpha_r$,
or whether considering the full sample, the reliable group, or the
reliable-$z_\mathrm{host}$ group. Subdividing the reliable-$z_\mathrm{host}$
group by redshift shows the same trend that was identified in
Figure~\ref{fig:model_parameters}: higher-redshift SNe have smaller values of
$\alpha$, on average, for the reliable-$z_\mathrm{host}$ group.

\section{Strong Priors}\label{sec:strong_priors}

\subsection{Fixing $\alpha = 2$}

In our previous effort to model the early evolution of SNe\,Ia, we adopted a
flexible model (hereafter the ``uninformative prior'') allowing $\alpha$ and
\tfl\ to simultaneously vary, despite theoretical \citep{Arnett82,Riess99a}
and observational \citep{Conley06,Hayden10,Ganeshalingam11,Gonzalez-Gaitan12}
evidence that $\alpha$ is consistent with $2$. Here, we alter the model by
fixing $\alpha_g = \alpha_r = 2$ (hereafter the ``$\alpha = 2$ prior''), and
explore how this decision changes the results described in the previous
sections. This decision is equivalent to placing an infinitely strong prior on
the value of $\alpha$.\footnote{Strictly enforcing $\alpha_g = \alpha_r$
imposes nonphysical structure on the models, as this condition effectively
implies that there is no change in the $g - r$ color during the initial rise
of the SN. This is clearly observed not to be the case in many SNe; see our 
\S\ref{sec:colors}, as well as Paper III in this series \citep{Bulla20}).}

\begin{figure}
    \centering
    \includegraphics[width=1\linewidth]{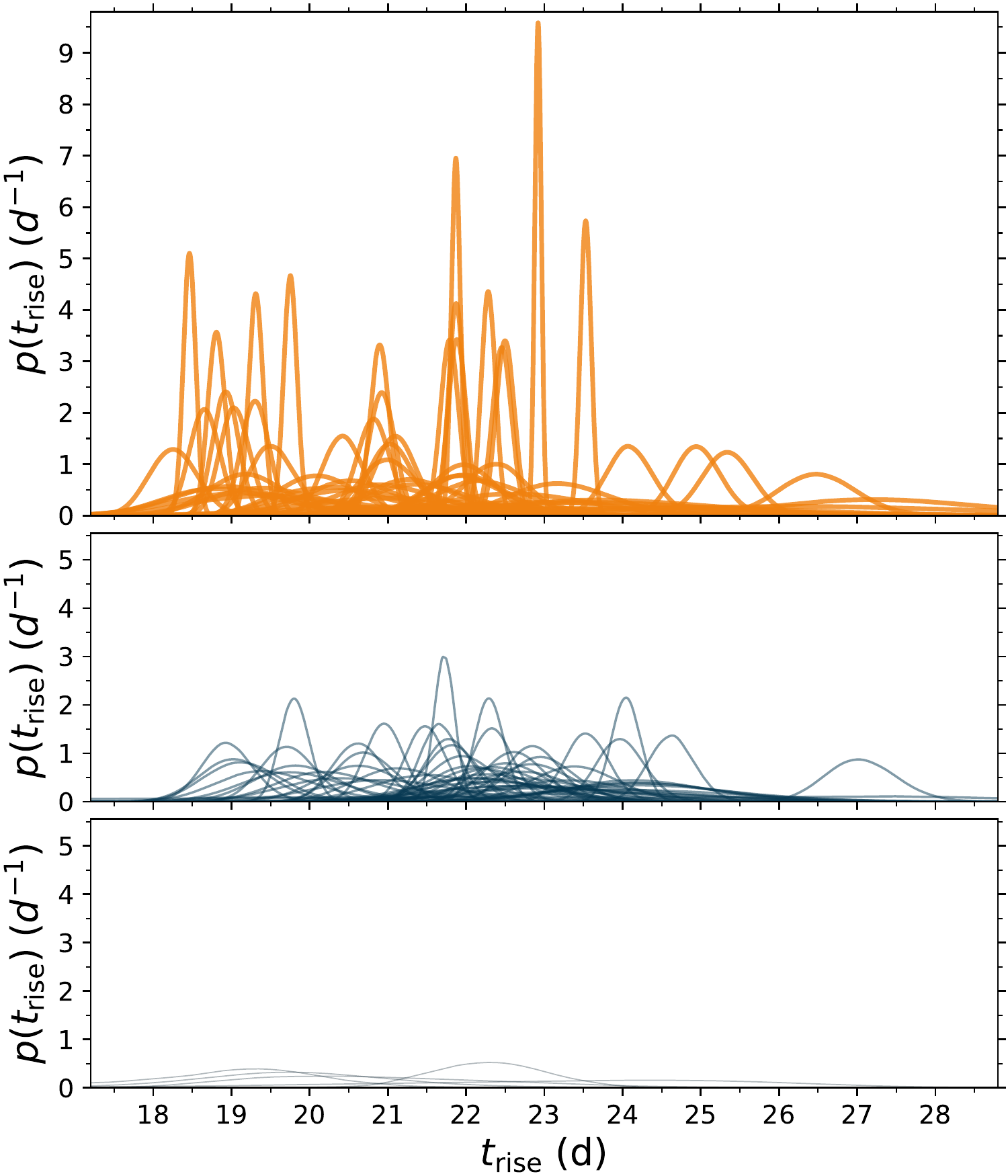}
    \caption{Same as Figure~\ref{fig:rise_time}, showing the resulting PDFs
    for the $\alpha = 2$ prior. Note that the definition for reliable model
    parameters with the $\alpha = 2$ prior is different from that described in
    \ref{sec:qa} (see text). As was the case when $\alpha$ is allowed to vary,
    there is no support for a single mean $t_\mathrm{rise}$ to describe every
    SN in the sample. The $\alpha = 2$ prior results in rise times that are
    $\sim$3\,d longer on average.}
    \label{fig:tsquared_rise}
\end{figure}

The distribution of rise-time PDFs using the $\alpha = 2$ prior is shown in
Figure~\ref{fig:tsquared_rise}, and reported in Table~\ref{tab:alpha2_rise}.
Adopting this strict prior significantly reduces the flexibility of the model.
One consequence of this choice is that visual inspection of the posterior
predictive flux values reveals that there are far fewer SNe with unreliable
model parameters. When using the $\alpha = 2$ prior, we only flag SNe with an
extrapolated flux using the maximum \textit{a posteriori} model parameters $<
0.9 f_\mathrm{max}$ at $T_{g,\mathrm{max}}$ as having unreliable model
parameters. Based on this criterion, only five SNe are identified as having
unreliable model parameters.

\begin{deluxetable}{llcccc}
\tabletypesize{\scriptsize}
\tablewidth{0pt}
\tablecaption{Ninety Percent Credible Region for $t_\mathrm{rise}$ ($\alpha = 2$ Prior)\label{tab:alpha2_rise}}
\tablehead{
\colhead{}
& \colhead{}
& \multicolumn{3}{c}{$t_\mathrm{rise}$\,(d)}
& \colhead{} \\
\cline{3-5}
\colhead{ZTF Name}
& \colhead{TNS Name}
& \colhead{5}
& \colhead{50}
& \colhead{95}
& \colhead{Reliable\tablenotemark{\scriptsize{a}}}
}
\startdata
ZTF18aailmnv & SN\,2018ebo & 21.01 & 22.22 & 23.65 & y \\
ZTF18aansqun & SN\,2018dyp & 17.82 & 19.22 & 20.88 & y \\
ZTF18aaoxryq & SN\,2018ert & 21.02 & 23.21 & 26.34 & y \\
ZTF18aapqwyv & SN\,2018bhc & 17.83 & 18.95 & 20.53 & y \\
ZTF18aapsedq & SN\,2018bgs & 21.71 & 22.38 & 23.03 & y \\
ZTF18aaqcozd & SN\,2018bjc & 18.12 & 19.59 & 21.52 & y \\
ZTF18aaqcqkv & SN\,2018lpc & 17.62 & 19.79 & 22.13 & n \\
ZTF18aaqcqvr & SN\,2018bvg & 21.18 & 21.98 & 22.91 & y \\
ZTF18aaqcugm & SN\,2018bhi & 19.06 & 19.52 & 20.04 & y \\
ZTF18aaqffyp & SN\,2018bhr & 17.91 & 23.36 & 26.79 & n \\
ZTF18aaqnrum & SN\,2018bhs & 16.47 & 22.55 & 25.51 & y \\
ZTF18aaqqoqs & SN\,2018cbh & 24.15 & 24.62 & 25.12 & y \\
ZTF18aarldnh & SN\,2018lpd & 19.32 & 20.57 & 22.04 & y \\
ZTF18aarqnje & SN\,2018bvd & 20.40 & 21.55 & 22.83 & y \\
ZTF18aasdted & SN\,2018big & 23.42 & 23.53 & 23.65 & y \\
\enddata
\tablecomments{
The table includes the $5^\mathrm{th}$, $50^\mathrm{th}$, 
and $95^\mathrm{th}$ percentiles for  
$t_\mathrm{rise}$ after adopting the $\alpha = 2$ prior (see text for further details).}

\tablenotetext{a}{Flag for SNe with reliable model parameters.
Note that the $\alpha = 2$ prior definition of reliable differs from that in 
\ref{sec:qa} (see text).}
(This table is available in its entirety in a machine-readable 
form in the online journal.)
\end{deluxetable}

\begin{figure}
    \centering
    \includegraphics[width=1\linewidth]{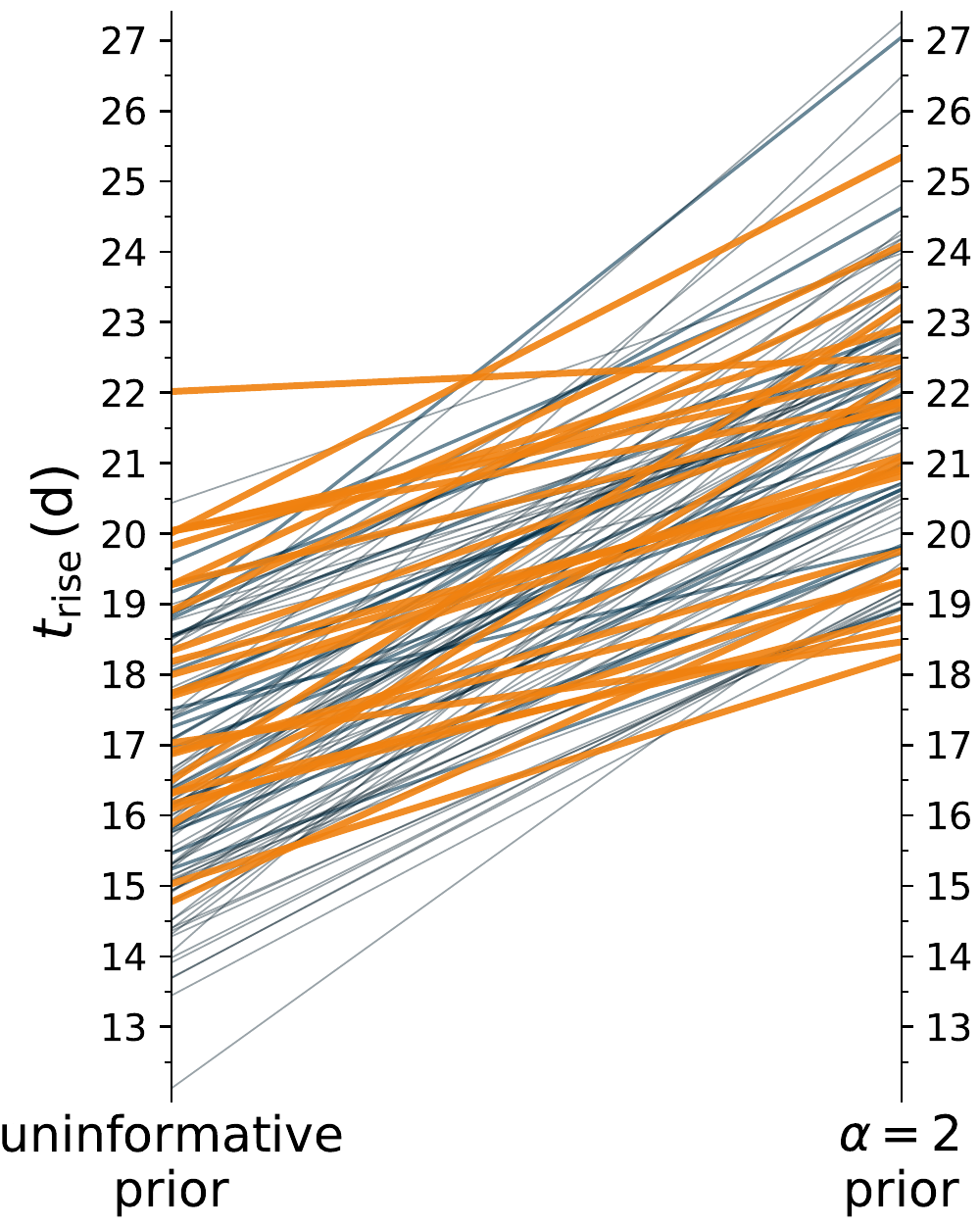}
    \caption{Comparison of the rise time inferred when adopting the
    uninformative prior and the $\alpha = 2$ prior for the 120 normal SNe in
    our sample. Adoption of the $\alpha = 2$ prior leads to larger values of
    \trise\ for every SN, with an average increase of $\sim$3\,d. Individual
    lines are colored orange, dark blue, and gray for the uninformative prior
    reliable-$z_\mathrm{host}$, reliable-$z_\mathrm{SN}$, and unreliable
    groups, respectively.}
    \label{fig:trise_prior}
\end{figure}

Figure~\ref{fig:trise_prior} shows how the inferred rise time changes
when adopting the $\alpha = 2$ prior instead of the uninformative prior for
every SN in our sample. The $\alpha = 2$ prior reduces the uncertainty on
\trise\ (compare Figures~\ref{fig:tsquared_rise} and~\ref{fig:rise_time}) and
increases the inferred rise time for each individual SN, with an average
increase of $\sim$3\,d in \trise.

Multiplying the individual likelihoods for \trise\ does not provide support
for a single mean rise time. Following the same approach described in
\S\ref{sec:mean_rise}, we find a population mean \trise\;$\approx 21.5$\,d,
with a corresponding population scatter of $\sim$2.0\,d for the $\alpha = 2$
prior (see Table~\ref{tab:mean_params}). Another consequence of adopting the
$\alpha = 2$ prior is that a small handful ($\sim$5--6) of SNe have rise times
consistent with 26\,d, which is considerably longer than the rise times
inferred in any previous study of normal SNe\,Ia.

\begin{figure*}
    \centering
    \includegraphics[width=6in]{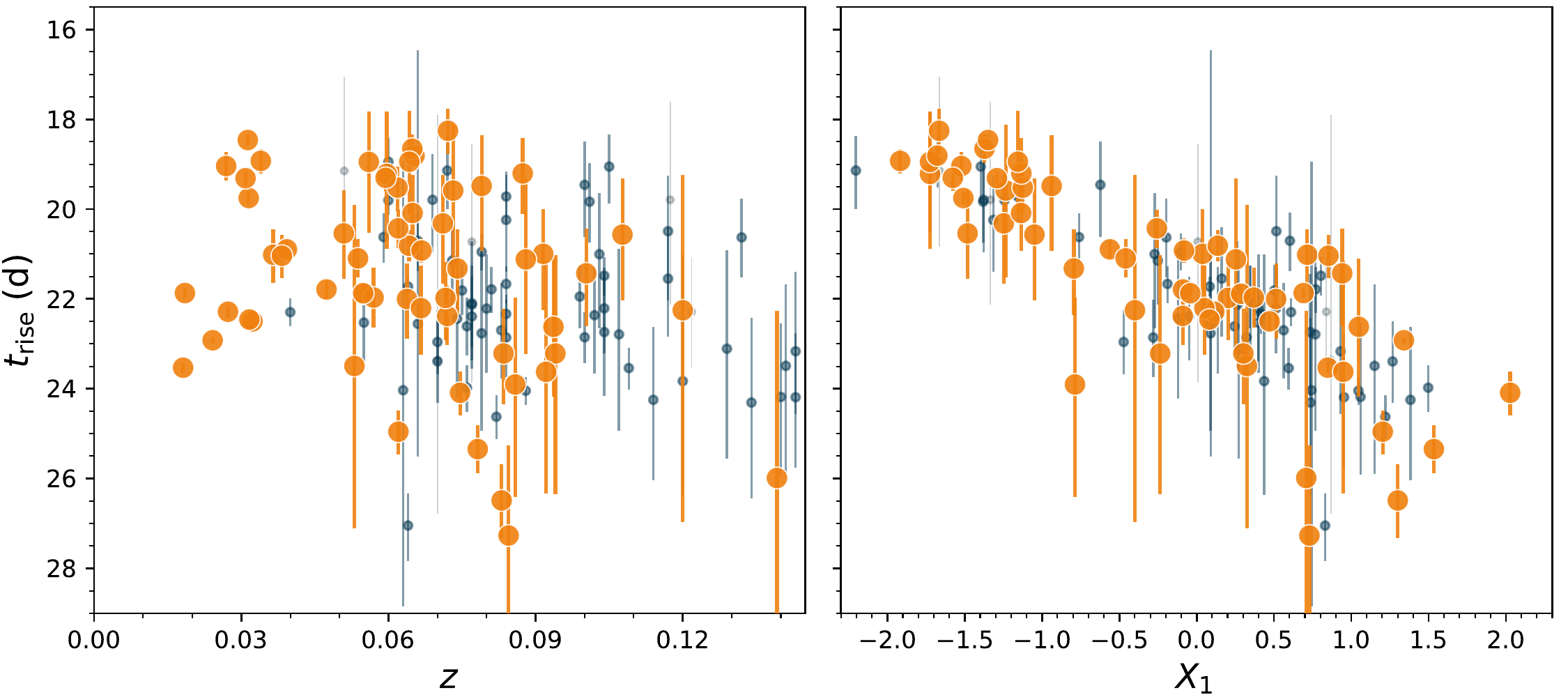}
    \caption{Correlation between \trise\ and redshift (left) and $x_1$ (right)
    when adopting the $\alpha = 2$ prior for the early emission from SNe\,Ia.
    Use of this strict prior removes the previously observed bias that
    resulted in shorter rise times being inferred at higher redshifts. One
    consequence of the removal of this bias is a reduction in the observed
    scatter between \trise\ and $x_1$.}
    \label{fig:tsquared_z_evolution}
\end{figure*}

Figure~\ref{fig:tsquared_z_evolution} shows \trise\ as a function of redshift
(left) and $x_1$ (right) when adopting the $\alpha = 2$ prior. The previously
observed trend where \trise\ decreases at increasing redshifts is no longer
seen. The model is, in effect, no longer flexible enough to systematically
adjust \tfl\ to be approximately equal to the epoch of first detection. The
removal of this particular bias provides a potential benefit of fixing $\alpha
= 2$.

Adopting the $\alpha = 2$ prior yields a significantly smaller scatter in the
correlation between $x_1$ and \trise, as shown in
Figures~\ref{fig:shape_correlations} and~\ref{fig:tsquared_z_evolution}. The
reduction in this scatter intuitively makes sense, given that it was due, at
least in part, to the redshift bias in measuring \trise\
(\S\ref{sec:redshift_correlations}). Reducing, or possibly fully removing,
that bias by adopting the $\alpha = 2$ prior allows a direct estimate of the
rise time from $x_1$ with a typical scatter $< 1$\,d. If relatively
high-precision measurements of \trise\ can be directly inferred from $x_1$, it
would dramatically increase the sample of SNe\,Ia with measured rise times, as
extremely early observations ($t < T_{B,\mathrm{max}}-10$\,d) would no longer
be required (see \S\ref{sec:x1_rise}).

\subsection{Model Selection}\label{sec:dic}

In adopting two very different priors that, in turn, produce significantly
different posteriors, we are naturally left with the question: which model is
better? To some extent, the answer to this question rests with each
individual, as the prior quantifies one's \textit{a priori} belief about the
model parameters. We posit that it is extremely unlikely that thermonuclear
explosions all identically produce $\alpha = 2$ across a multitude of filters.
Adopting $\alpha = 2$ is therefore very likely an overconfident position that
produces slightly biased inferences.

Alternatively, we can address the question of which model is best via the use
of model selection techniques based on information criteria. The
uninformative prior includes more model parameters as $\alpha$ is allowed to
vary. For individual SNe, we can compare the trade-off between increasing the
model complexity, relative to the $\alpha = 2$ prior, and the overall
improvement in the fit to the data in order to determine which model is
superior. Following \citet{Spiegelhalter02}, we define the deviance $D$ as
$$D(\theta) = -2 \ln \left(p(x\mid \theta)\right) + C,$$
where $\theta$ represents the model parameters, $x$ represents the
observations, $p(x\mid \theta)$ is the likelihood, and $C$ is a constant that
will drop out following model comparison. From here, the effective
number\footnote{Many model selection techniques based on information criteria,
such as the Akaike information criterion \citep[AIC;][]{Akaike74} or
``Bayesian'' information criterion \citep[BIC;][]{Schwarz78}, include a point
estimate of the maximized likelihood and then subtract a penalty term related
to the total number of parameters in the model. These methods aim to balance
the goodness of fit while regulating the overall model complexity. The AIC and
BIC cannot, however, easily be applied to the models adopted here, as the use
of an informative prior distribution reduces the amount of overfitting and
produces an ``effective'' number of parameters that is smaller than the number
of variables in the parameterized model \citep{Gelman14}. We adopt the DIC, as
opposed to the AIC or BIC, because it effectively marginalizes over the
nuisance parameters and applies to the full set of posterior predictions.} of
model parameters $p_D$ can be calculated as:
$$p_D = \langle D(\theta) \rangle - D(\langle \theta \rangle),$$
where $\langle D(\theta) \rangle$ is the mean posterior value of the deviance,
and $D(\langle \theta \rangle)$ is the deviance of the mean posterior model
parameters. We then define the deviation information criterion (DIC) as:
$$\mathrm{DIC} = p_D + \langle D(\theta) \rangle.$$
Smaller values of  the DIC are  preferred to larger values.

Following \citet{Jeffreys61}, we consider SNe with
$$\exp\left(\frac{\mathrm{DIC}_{\alpha2} - \mathrm{DIC}_\mathrm{flat}}{2}\right) \ge 30,$$
where $\mathrm{DIC}_{\alpha2}$ is the DIC for the $\alpha = 2$ prior and
$\mathrm{DIC}_\mathrm{flat}$ is the DIC for the uninformative prior, to
exhibit a very strong preference for the uninformative prior. Of the 127 SNe
in our sample, including the 7 SNe that are not considered normal SNe\,Ia,
only 29 show a strong preference for the uninformative prior. Of these 29 SNe,
16 belong to the unreliable group. Visual inspection of these 16 confirms that
these SNe have very few detections after \tfl. In these cases, the data are
fit extremely well with very small values of $\alpha$ (see, e.g.,
Figure~\ref{fig:biggap_lc}). The remaining 13 SNe are at low $z$, with few, if
any, gaps in observational coverage.

Thus, the $\alpha = 2$ prior should be used to estimate \tfl\ for all but 13
SNe in our sample. For these 13, the uninformative prior provides a better
estimate of \tfl, according to the DIC. This combination of results is how we
define the distribution of \tfl\ in Paper III of this series \citep{Bulla20}.

\section{A Volume-limited Sample of Normal SNe\,Ia}\label{sec:volume_limited}

The ZTF sample of SNe\,Ia is clearly biased due to Malmquist selection effects
(see \citealt{Yao19}), and as such, the population results discussed above are
also correspondingly biased. We can, however, approximate a volume-limited
subset of \textit{normal} SNe\,Ia. A full study of the completeness of the ZTF
SNe\,Ia sample is beyond the scope of this paper and will be discussed in a
future study (J.~Nordin et al. 2020, in preparation).

The selection criteria presented in Paper I removes SNe\,Ia from the sample if
they lack a \gztf\ detection $> 10$\,d prior to \tbmax\ \citep{Yao19}. By
construction, the intrinsically faintest normal SNe\,Ia in the ZTF sample have
$x_1 \approx -2$, with $M_g \approx -17$\,mag at $t \approx -10$\,d. With a
typical limiting magnitude of $g_\mathrm{ztf} \approx 20.0$\,mag during bright
time \citep{Bellm19}, the ZTF high-cadence survey should be complete to all
$x_1 \approx -2$ and brighter SNe to a distance modulus $\mu \approx 37$\,mag.
For our adopted cosmology, this distance corresponds to a redshift $z \approx
0.0585$. Thus, the 28 normal SNe\,Ia with $z < 0.06$ should comprise a
volume-limited subset of our sample.

For the uninformative prior, 16 of the 28 low-redshift SNe have reliable model
parameters, and 15 of those 16 have known host redshifts. Using the same
procedure as \S\ref{sec:mean_rise}, we estimate a weighted mean rise time of
$\sim$18.9\,d when considering the volume complete subset ($z < 0.06$)
of our sample (see Table~\ref{tab:mean_params} for the mean values discussed
here and in the remainder of this section). For these SNe, we also find mean
values of $\sim$2.13 and $\sim$2.01 for $\alpha_g$ and $\alpha_r$,
respectively.

For the $\alpha = 2$ prior, 27 of the 28 low-redshift SNe have reliable model
parameters, and 24 of those 27 have known host redshifts. For this prior, we
estimate a weighted mean rise time of $\sim$21.7\,d.

If we instead use the results from the $\alpha = 2$ prior, unless the DIC
provides very strong evidence for the uninformative prior, as suggested at the
end of \S\ref{sec:dic}, then we find a mean rise time of $\sim$19.9\,d
for the volume-limited sample. It makes sense that this mean is nearly 2\,d
shorter than the mean for the $\alpha=2$ prior, because the SNe for which the
DIC prefers the uninformative prior provide the lowest variance estimates of
\trise.

Finally, if examine only those SNe for which the DIC prefers the uninformative
prior, of which there are only nine normal SNe with $z < 0.06$ (all of which
have known $z_\mathrm{host}$) in our entire sample, we find a mean rise time
of $19.20 \pm 0.06$. For this same subset, we find mean values of $\alpha_g$
and $\alpha_r$ of $2.14 \pm 0.03$ and $2.02 \pm 0.03$, respectively.

Given the bias identified in \S\ref{sec:redshift_correlations}, it is not
surprising that a volume-limited sample of SNe has larger estimates for the
mean values of \trise\ and $\alpha$, when using the uninformative prior. For
the $\alpha = 2$ prior, on the other hand, the volume-limited sample produces
very similar estimates for the mean \trise\ ($< 1\%$ difference) and the full
sample. This provides additional evidence that the adoption of a strong prior
can negate the redshift bias highlighted in \S\ref{sec:redshift_correlations}.

\section{Discussion}

\subsection{SNe\,Ia Rise Times}

In the analysis above, we provide multiple measurements of the rise time of
SNe Ia following the adoption of different priors. Within the literature,
there are at least a half dozen entirely different methods that have been
employed to answer precisely the same question. This naturally raises the
question -- which method is best? Answering that question in turn raises an
important offshoot as well -- is the method cheap to implement (i.e., does it
provide reliable inference in the limit of poor sampling or low S/N)?

\subsubsection{\texttt{SALT2} $x_1$ as a Proxy for \trise}\label{sec:x1_rise}

Estimating the rise times of SNe\,Ia using only observations around maximum
brightness would be an ideal approach. This approach would maximize the sample
size from flux-limited surveys, and Figure~\ref{fig:tsquared_z_evolution}
suggests it may be feasible given the correlation between the \texttt{SALT2}
$x_1$ parameter and \trise\ (measured using the $\alpha = 2$
prior).\footnote{Other shape parameters, such as the stretch, $s$, or distance
from a fiducial template, $\Delta$, may work in place of $x_1$.} Even in the
limit of only one or a few observations on the rise, \texttt{SALT2} can still
measure $x_1$ (e.g., \citealt{Scolnic18a}). Therefore, the correlation between
$x_1$ and \trise\ eliminates the need for high-cadence observations to yield
early ($> 10$\,d prior to \tbmax) discoveries, enabling a more economical
method to estimate \trise\ relative to the methods described above.

For the volume-limited sample (see \S\ref{sec:volume_limited}) of normal SNe
Ia with known host galaxy redshifts and reliable model parameters, we estimate
the relation between \trise\ and $x_1$ via a maximum-likelihood linear fit
that accounts for the uncertainties on both \trise\ and $x_1$ (see
\citealt{Hogg10}). From this fit, we find
\begin{equation}
    t_\mathrm{rise} = (21.41 \pm 0.03) + (1.62 \pm 0.03)x_1\,\mathrm{d}.
    \label{eqn:rise_x1}
\end{equation} 
The residual scatter about this relation, as estimated by the sample standard
deviation, is 0.81\,d. We find the relation does not significantly change when
including SNe with unknown host-galaxy redshifts or unreliable model
parameters (though the scatter increases to $\sim$1.2\,d when including $z \ge
0.06$ SNe in the fit). Thus, if one assumes $\alpha = 2$, then \texttt{SALT2}
can be used to estimate \trise\ with a typical uncertainty of $\sim$0.8\,d.
This scatter is only slightly worse than the median uncertainty on \trise,
$\sim$0.5\,d, for individual SNe when adopting the $\alpha = 2$ prior (see
\S\ref{sec:strong_priors}). Furthermore, given that an $x_1 = 0$ SN is
supposed to represent a ``mean'' SN Ia, Equation~\ref{eqn:rise_x1} suggests
that the mean rise time of SNe\,Ia is $\sim$21.4\,d.

If we repeat the same exercise using uninformative prior rise times for the
volume limited sample, we find that the typical scatter about the linear
\trise-$x_1$ relation is $\sim$1.7\,d and $\sim$1.4\,d for the full sample and
reliable group, respectively. The rise time of a mean SN according to this
relation is $\sim$18.4\,d, however, we caution that some individual
rise time measurements for this prior may be underestimated as discussed in
\S\ref{sec:redshift_correlations}.

\subsubsection{Precise Estimates of \trise\ from Early Observations}

While the \trise--$x_1$ relation provides a relatively cheap method to infer
the rise time of normal SNe\,Ia, a significant advantage of early observations
is that they can provide far more precise estimates of \trise, especially in
the limit of high S/N. For the $\alpha = 2$ prior, there are 14 SNe with a half
68\% credible region that is $< 3$\,hr. For the uninformative prior, this
number drops to two SNe. In either case, these measurements provide far more
precision than possible from extrapolations based on SN Ia shape parameters
(such as $x_1$).

While the methods adopted in this paper provide higher precision, it is
impossible that they are both accurate. The median difference in the inferred
\trise\ from the $\alpha = 2$ and uninformative priors for individual SNe is
4.9\,d. Even the volume-limited subset of SNe with reliable model parameters
from the uninformative prior (15 total SNe) have a median difference of 2.9\,d
in \trise\ for the two priors. The systematic effect identified in
\S\ref{sec:redshift_correlations} suggests that the uninformative prior does
not provide accurate estimates of \trise\ for higher-$z$ SNe. The $\alpha = 2$
prior, on the other hand, explicitly assumes that there is no change in the
early optical color of SNe\,Ia. Many SNe with early observations clearly
invalidate this particular assumption, raising the possibility that neither
method is accurate.

An SN cannot be detected until it has exploded, and thus the epoch of discovery
provides a lower limit on \trise. Between PTF/iPTF \citep{Papadogiannakis19}
and ZTF \citep{Yao19}, there are $\sim$20 SNe\,Ia that are detected at least
18\,d before \tbmax, with a few detections as early as 21\,d before \tbmax. If
the uninformative prior is accurate, then each of these SNe would represent an
incredibly lucky set of circumstances: (i) they each have longer rise times
than average ($\sim$18\,d; see \S\ref{sec:volume_limited}), and (ii) they were
all discovered more or less immediately after \tfl. A more probable
explanation is that the mean rise time is $> 18$\,d, in which case the $\alpha
= 2$ prior may provide a more accurate inference of the rise time (though
again we caution that these estimates may also be inaccurate). The fact that
each of the four SNe with $z < 0.03$, which should have the least biased rise
time estimates (see \S\ref{sec:redshift_correlations}), have \trise$ > 18$\,d,
further supports this claim.

\subsubsection{Comparison to the Literature}

Several studies in the literature have attempted to measure the mean rise time
of SNe\,Ia. Here, we compare our work to previous results. This exercise is
somewhat fraught with difficulty, in the sense that each study incorporates
slight differences in implementation, which in turn makes comparisons
challenging. Furthermore, these studies are typically conducted with different
filter sets and over a wide range of redshifts, which may introduce biases
that are difficult to quantify across studies (as discussed above, $K$
corrections are highly uncertain at very early epochs). Finally, the quality
of the data in each of these studies is vastly different. For example, in
\citet{Riess99a} there are only six SNe (and 10 total $B$-band observations)
observed at phases $\le -15$\,d, while our study includes 31 SNe
\textit{discovered $> 15$\,d before \tbmax} \citep{Yao19}. As we proceed with
our cross-study comparison, we exclude rise time estimates for individual SNe
and instead focus on studies with relatively large samples ($\gtrsim 10$
normal SNe\,Ia).

As outlined in the \nameref{sec:intro}, there are broadly two different
methods to measure the mean rise time of SNe\,Ia. The first uses the
well-established luminosity--decline relation for SNe\,Ia \citep{Phillips93}
to ``shape correct'' the SN light curves prior to fitting for the rise time.
Thus, individual light curves are stretched by some empirically measured
factor, and the mean rise time represents a normal SN Ia \textit{after shape
correction} \citep[e.g.,][]
{Riess99a,Aldering00,Conley06,Hayden10,Ganeshalingam11,Gonzalez-Gaitan12}. The
second method measures the rise time of each SN Ia within a sample, and then
takes the mean of this distribution. These two methods are not equivalent, and
therefore are likely to produce different results. If, for instance, a
flux-limited survey finds more high-luminosity, slower-declining SNe than
low-luminosity, faster-declining events, then the population mean will produce
longer rise times than the shape-corrected mean.

Using different data sets obtained in very different redshift regimes,
\citet{Riess99a}, \citet{Aldering00}, and \citet{Conley06} estimate consistent
values of the shape-corrected mean \trise$ \approx 19.5$\,d. Each of these
studies fixes $\alpha = 2$ when fitting for the rise time. This
estimate is similar to, though slightly longer, than our estimate for the mean
rise time in the reliable, volume-limited subset of our sample (see
\S\ref{sec:volume_limited}). Later studies by \citet{Hayden10},
\citet{Ganeshalingam11}, and \citet{Gonzalez-Gaitan12} also provide estimates
of the mean shape-corrected rise time and find smaller values of
$\sim$17.0--18.0\,d. As noted by \citeauthor{Hayden10}, these methods are
highly dependent on the template light curve used to stretch the individual
SNe, and differences in the templates used by each study may explain the
dissensus between their findings.

The approach employed in \citet{Firth15}, \citet{Zheng17a}, and
\citet{Papadogiannakis19} is more similar to the one adopted here. Each of
these studies estimates \trise\ for individual SNe and then calculates the
population mean. If the samples differ between any of these studies--and aside
from 11 SNe that are included in both \citet{Papadogiannakis19} and
\citet{Firth15}, there is no overlap between any of those studies or this
one--then it should be expected that the population mean rise time estimates
will differ. Furthermore, the \trise\ estimates in \citet{Papadogiannakis19}
and \citet{Firth15} are not relative to \tbmax, and it is known that the rise
time varies as a function of wavelength (e.g., \citealt{Ganeshalingam11}).
Taken together, this confluence of factors makes it difficult to compare
results between these studies, which we nevertheless do below.

In \citet{Zheng17}, a semi-analytical, six-parameter, broken power-law model
is introduced to describe the optical evolution of SNe\,Ia. This model has a
distinct advantage over the methods employed here, in that an artificial cutoff
does not need to be applied in flux space (see \S\ref{sec:model}), though a
post-peak cut must be applied as the model cannot reproduce the evolution of
SNe into the nebular phase. A downside of this approach is that there are
large degeneracies between the different model parameters, meaning it is
difficult to find numerically stable solutions without fixing individual
parameters to a single value \citep{Zheng17a}. For a sample of 56
well-observed low-$z$ SNe, this method produces a mean rise time of 16.0\,d
\citep{Zheng17a}, while the same technique applied to SNe\,Ia from PTF/iPTF
finds a mean rise time of 16.8\,d \citep{Papadogiannakis19}. These estimates
are considerably lower than the ones presented here, and are almost certainly
underestimates of the true mean rise time based on the large number of SNe
with detections $>$16\,d before peak \citep{Papadogiannakis19, Yao19}. Indeed,
inspection of Figure~1 in \citet{Zheng17a} shows that the six-parameter model
underestimates the flux at the very earliest epochs and underestimates \trise\
as a result.

The closest comparison to the methods used in this study can be drawn from
\citet{Firth15}. Using a sample of 18 SNe discovered by PTF and the La Silla
Quest (LSQ) survey, \citeauthor{Firth15} fit a model similar to
Equation~\ref{eqn:flux_model}, in that \tfl\ and $\alpha$ are allowed to
simultaneously vary. From these fits, they estimate a mean population \trise$
= 18.98 \pm 0.54$\,d, which is consistent with our estimate of the rise time
for the volume-complete $z_\mathrm{host}$ sample, $\sim$18.9\,d. Contrary to
this study, they find shorter rise times, and a mean of $\sim$17.9\,d, when
fixing $\alpha = 2$ (this is likely explained by their adopted fit procedure;
see \S\ref{sec:fireball_discussion}).

\subsection{The Expanding Fireball Model}\label{sec:fireball_discussion}

The expanding fireball model (see \S\ref{sec:model}) is remarkable in its
simplicity. Its two underlying assumptions, that the photospheric velocity and
temperature of the ejecta are approximately constant during the early
evolution of the SN, are clearly over-simplifications; \citet{Parrent12} shows
that the photospheric velocity declines by at least 33\% in the $\sim$5\,d
after explosion. Despite these simplifications, numerous studies have found
that $\alpha$ is consistent with 2 (e.g.,
\citealt{Conley06,Hayden10,Ganeshalingam11,Gonzalez-Gaitan12,Zheng17a}).

Based on the volume-limited subset of normal SNe\,Ia with reliable model
parameters, we find a population mean $\alpha_r = 2.01 \pm 0.03$, which is
consistent with the expanding fireball model. For $\alpha_g$, on the other
hand, we find a population mean of $2.13 \pm 0.03$, which is only marginally
consistent with 2. As previously noted, a power-law index of 2 in every
optical filter would stand in contrast to observations, as it would imply that
SNe\,Ia have no color evolution shortly after explosion.

Furthermore, there are individual normal SNe\,Ia for which the expanding
fireball model does not apply. There are several examples within the
literature (e.g.,
\citealt{Zheng13,Zheng14,Goobar15,Miller18,Shappee19,Dimitriadis19}), and
within this study, several low-$z$ SNe are inconsistent with the expanding
fireball model:
\begin{itemize}
\item ZTF18aasdted (SN\,2018big; $z \approx 0.018$, $\alpha_r \approx 1.4$),
\item ZTF18abauprj (SN\,2018cnw; $z \approx 0.024$, $\alpha_r \approx 2.2$),
\item ZTF18abcflnz (SN\,2018cuw; $z \approx 0.027$, $\alpha_r \approx 2.4$),
\item ZTF18abfhryc (SN\,2018dhw; $z \approx 0.032$, $\alpha_r \approx 3.3$), 
\item ZTF18abuqugw (SN\,2018geo; $z\,\approx\,0.031$, $\alpha_r\approx2.7$).
\end{itemize}
In each of these cases, the DIC clearly prefers $\alpha \neq 2$.

Given that many of the very best-observed, low-redshift SNe are incompatible
with $\alpha = 2$, and that the mean $\alpha_g > 2$ in the ZTF sample, it is
clear that the expanding fireball model does not adequately reproduce the
observed diversity of SNe\,Ia. Nevertheless, according to the
DIC, $\alpha=2$ provides a reasonable proxy for the early evolution of the
majority of normal SNe\,Ia (at the quality of ZTF high-cadence observations).
This is either telling us that individual SNe exhibiting significant
departures from $\alpha = 2$ are atypical--an interpretation adopted in
\citealt{Hosseinzadeh17,Miller18,Dimitriadis19} and elsewhere--or that, for
the vast majority of SNe, the observations are not of high enough quality to
conclusively show $\alpha \neq 2$. Distinguishing between these two
possibilities requires larger volume-limited samples.

Moving forward, it may be that the most appropriate prior for fitting the
early evolution of SNe\,Ia is to adopt a Gaussian centered at 2 for $\alpha$
in the redder filters, while also placing a prior on the difference in
$\alpha$ across different filters (for ZTF $\alpha_r - \alpha_g \approx -0.18$
based on \S\ref{sec:colors}). More testing and observations, especially of
low-$z$ SNe, and testing, are needed to confirm whether or not such priors are
in fact appropriate.

Finally, we note that the analysis in \citet{Firth15} finds a mean value of
$\alpha = 2.44 \pm 0.13$, which is not consistent with 2. This result can be
understood in the context of the \citet{Firth15} fitting procedure, whereby an
initial estimate of \tfl\ is made by fixing $\alpha = 2$. Only observations
obtained 2 days before and after this initial \tfl\ estimate are included in
the final model fit (i.e., the entire baseline of nondetections is not used,
as is done in this study). Truncating the baseline biases the model to longer
rise times (as is observed in \citet{Firth15}), and, as shown in
Figures~\ref{fig:corner_good}--\ref{fig:corner_bad}, longer rise times require
larger values of $\alpha$ when adopting a simple power-law model (as is done
here and in \citet{Firth15}).

The reason it is critical to test the expanding fireball model is that robust
measurements of $\alpha$ can distinguish between different explosion
scenarios. For example, the delayed-detonation models presented in
\citet{Blondin13}, which provide a good match to SNe\,Ia at maximum light,
systematically overestimate the power-law index at early times (with typical
values of $\alpha \approx 7$; see Figure~1 in \citet{Dessart14}). This led
\citet{Dessart14} to alternatively consider pulsational-delayed detonation
models, which do result in a smaller power-law index ($\alpha \approx 3$),
though those results are still incompatible with what we find
here.\footnote{The range of $\alpha$ values reported in \citet{Dessart14} is
fit to the first $\sim$3\,d after explosion. Fitting all observations with
$f_\mathrm{obs} \leq 0.4 f_\mathrm{max}$, as is done in this study, would
reduce the inferred values of $\alpha$ in \citet{Dessart14}.} In
\citet{Noebauer17}, the early evolution of various explosion models does not
follow an exact power law. They find an almost power-law evolution for pure
deflagration models, which may explain the origin of SN\,2002cx-like SNe. For
pure deflagrations, \citet{Noebauer17} find $\alpha < 2$, which qualitatively
agrees with our results for ZTF18abclfee (SN\,2018cxk), where $\alpha_r
\approx 1$ (see \ref{sec:rare}). The models presented in \citet{Magee20},
which examine the evolution of SNe with different $^{56}$Ni distributions,
provide good qualitative agreement to what we find for ZTF SNe.
\citet{Magee20} find that the rising power-law index is larger in the $B$ band
than the $R$ band (similar to what we see in \gztf\ and \rztf), and that the
mean value of these distributions is $\sim$2. As suggested in \citet{Magee20},
it may be the case that the vast majority of the differences observed in the
early ZTF light curves can be explained via variations in the $^{56}$Ni mixing
in the SN ejecta. Future modeling will test this possibility (M. Deckers et
al.\ 2020, in preparation).

\section{Conclusions}

In this paper, we have presented an analysis of the initial evolution and rise
times of 127 ZTF-discovered SNe\,Ia with early observations; see \citet{Yao19}
for details on how the sample was selected. These SNe were observed as part of
the ZTF high-cadence extragalactic experiment, which obtained three \gztf\ and
three \rztf\ observations every night the telescope was open. A key
distinction of this data set, in contrast to many previous studies, is the
large number of observations taken prior to the epoch of discovery, which
meaningfully constrains the behavior of the SN at very early times (see
Appendix~\ref{sec:pre_explosion}). The uniformity, size, and observational
duty cycle of this data set are truly unique, making this sample of ZTF SNe
the premier data set for studying the early evolution of thermonuclear SNe.

We model the emission from these SNe as a power law in time $t$, whereby the
flux $f \propto (t - t_\mathrm{fl})^\alpha$, where \tfl\ is the time of first
light, and $\alpha$ is the power-law index. By simultaneously fitting
observations in the \gztf\ and \rztf\ filters, we are able to place stronger
constraints on \tfl\ than would be possible with observations in a single
filter. While many previous studies have fixed $\alpha = 2$, following the
simple expanding fireball model (e.g., \citealt{Riess99a}), we have instead
allowed $\alpha$ to vary, as there are recent examples of SNe\,Ia where
$\alpha$ clearly is not equal to 2 (e.g.,
\citealt{Zheng13,Zheng14,Goobar15,Miller18,Shappee19,Dimitriadis19}). While
the population mean value of $\alpha$ tends toward 2, there are several
individual SNe featuring an early evolution that deviates from an $\alpha = 2$
power law, justifying our model parameterization.

As might be expected, we find that our ability to constrain the model
parameters is highly dependent on the quality of the data. SNe\,Ia at low
redshifts that lack significant gaps in observational coverage are better
constrained than their high-redshift counterparts or events with large
temporal gaps. We identify those SNe with reliable model parameters under the
reasonable assumption that models of the initial flux evolution should
overestimate the flux at peak brightness. Following this procedure, we find
that 51 of the SNe have reliable model parameters. We focus our analysis on
these events.

For the subset of normal SNe with reliable model parameters, we estimate a
population mean \trise\;$\approx 18.5$\,d, with a sample standard deviation of
$\sim$1.6\,d. For individual SNe, the range of rise times extends from
$\sim$15 to 22\,d. We have additionally identified a systematic in the
parameter estimation for models that simultaneously vary \tfl\ and $\alpha$.
Namely, for flux-limited surveys, the model constraints on \trise\ will be
systematically underestimated for the higher redshift SNe in the sample. If we
restrict the sample to a volume-limited subset of SNe ($z < 0.06$), where this
bias may still be present but probably less prevalent, we estimate a mean
population rise time of $\sim$18.9\,d.

Normal SNe\,Ia have a population mean $\alpha_g \approx 2.1$ and a population
mean $\alpha_r \approx 2.0$, with a population standard deviation $\sim$0.5
for both parameters. While the mean value for our sample of SNe tends toward
2, we observe a range in $\alpha$ extending from $\sim$1.0 to 3.5. For both
\trise\ and $\alpha$, there is no single value that is consistent with all the
SNe in our sample. Interestingly, we find that nearly all SNe are consistent
with a single value of $\alpha_r - \alpha_g$, which describes the initial
\gztf$ - $\rztf\ color evolution of SNe\,Ia. The data show a mean value of
$\alpha_r - \alpha_g \approx -0.18$, meaning the optical colors of most
SNe\,Ia evolve to the blue with comparable magnitudes over a similar
timescale. This could be a sign that the degree of $^{56}$Ni-mixing in the SN
ejecta is very similar for the majority of SNe\,Ia (e.g.,
\citealt{Piro16,Magee18,Magee20}).

We find that the rise time is correlated with the light-curve shape of the SN,
in the sense that high-luminosity, slowly declining SNe have longer rise
times. This finding is consistent with many previous studies.

Given the large number of SNe with unreliable model parameters, in addition to
the observed bias in the measurement of \trise\ for high-$z$ SNe, we also
consider how the model parameter estimates change with strong priors. In
particular, we adopt $\alpha_g = \alpha_r = 2$, enforcing the expanding
fireball hypothesis on the data. Strictly speaking, this prior means that the
early colors of SNe Ia do not change, which we empirically know is not the
case. Nevertheless, a nearly constant temperature is one of the assumptions of
the fireball model, and thus we proceed.

Under the $\alpha = 2$ prior, we find that far more SNe have reliable \trise\
estimates. For the typical SN in our sample, fixing $\alpha = 2$ results in an
increase in \trise\ by a few days. We estimate a population mean \trise\;$
\approx 21.6$\,d when adopting the $\alpha = 2$ prior. One consequence of
adopting this prior is that it significantly reduces the previously observed
bias where high-$z$ SNe are inferred to have shorter rise times. The use of
this prior also reduces the scatter in the $x_1$--\trise\ relation, and we
find that with \texttt{SALT2}, via the measurement of $x_1$, it is possible to
estimate \trise\ with a typical scatter of $\sim$0.81\,d, even if there are no
early-time observations available. We also find that, for the vast majority of
the SNe in our sample (all but 13 events) there is at best only weak evidence
that the $\alpha \ne 2$ model is preferred to a model with $\alpha_g =
\alpha_r = 2$ according to the DIC.

While we have primarily focused on the properties and evolution of normal SNe
Ia, there are seven SNe in our sample that cannot be categorized as normal (see
\citealt{Yao19}). In \ref{sec:rare}, we find that the rise
times of Ia-CSM SNe and SC explosions are longer than those of normal SNe\,Ia,
in accordance with previous studies. We highlight our observations of
ZTF18abclfee (SN\,2018cxk), an SN\,2002cx-like SN with exquisite observational
coverage in the time before explosion. We estimate \tfl\ to within
$\sim$8\,hr for ZTF18abclfee, making our measurement the most precise estimate
of \trise\ for any 02cx-like SN to date. ZTF18abclfee took $\sim$10\,d to
reach peak brightness, roughly 5\,d less than SN\,2005hk, another 02cx-like
event with a well-constrained rise time.

This study has important implications for future efforts to characterize the
rise times of SNe\,Ia. We have found that, for all but the best-observed,
highest S/N events, a generic power-law model where $\alpha$ is allowed to
vary does not place meaningful constraints on the rise time--or worse, in the
case of higher-$z$ events, it produces a biased estimate. If this were the end
of the story, it would be particularly bad news for LSST, which will typically
have gaps of several days in its observational cadence \citep{Ivezic19}. With
only Equation~\ref{eqn:flux_model} at our disposal, we would rarely be able to
infer \trise\ for LSST SNe. As we demonstrated with the $\alpha = 2$ prior, in
the limit of low-quality data, the application of a strong prior can
significantly improve the final inference. Our current challenge is to develop
an empirically motivated prior for the model parameters. This provides a
strong justification for the concurrent operation of LSST and small-aperture,
high-cadence experiments, such as ZTF and the planned ZTF-II. These smaller,
more focused, missions can provide exquisite observations of a select handful
of SNe that can be used to drive the priors in our inference. While there have
been thousands of SNe\,Ia studied to date (e.g., \citealt{Jones17}), more
examples are indeed still needed: there are only four normal, $z < 0.03$
SNe\,Ia in our sample, and these four are nearly as valuable as the entire
remainder of the sample for establishing the diversity of SNe\,Ia. As low-$z$,
high-cadence surveys improve our understanding of these priors, we can combine
that knowledge with the hitherto unimaginable statistical samples from LSST
($\sim$millions of SNe), to better understand the early evolution and rise
time distribution of Type Ia SNe.

\acknowledgements

The authors would like to thank the anonymous referee for helpful
comments that have improved this paper. We thank M.~Magee for sharing details
about the rise times of SNe\,Ia models. A.A.M.~would like to thank
E.~A.~Chase, M.~Zevin, and C.~P.~L.~Berry for useful discussions on KDEs and
PDFs. We also appreciate D.~Goldstein's suggestions regarding \texttt{SALT2}
as a proxy for rise time. Y.~Yang, J.~Nordin, R.~Biswas, and J.~Sollerman
provided detailed comments on an early draft that improved this manuscript.

A.A.M.~is funded by the Large Synoptic Survey Telescope Corporation, the
Brinson Foundation, and the Moore Foundation in support of the LSSTC Data
Science Fellowship Program; he also receives support as a CIERA Fellow by the
CIERA Postdoctoral Fellowship Program (Center for Interdisciplinary
Exploration and Research in Astrophysics, Northwestern University). Y.~Y.,
U.~C.~F., and S.~R.~K.~thank the Heising-Simons Foundation for supporting ZTF
research (\#2018-0907). This research was supported in part through the
computational resources and staff contributions provided for the Quest
high-performance computing facility at Northwestern University which is
jointly supported by the Office of the Provost, the Office for Research, and
Northwestern University Information Technology. This work was supported in
part by the GROWTH project funded by the National Science Foundation under
Grant No.~1545949.

This work is based on observations obtained with the Samuel Oschin Telescope
48 inch and the 60 inch Telescope at the Palomar Observatory as part of the
Zwicky Transient Facility project. ZTF is supported by the National Science
Foundation under Grant No. AST-1440341 and a collaboration including Caltech,
IPAC, the Weizmann Institute for Science, the Oskar Klein Center at Stockholm
University, the University of Maryland, the University of Washington,
Deutsches Elektronen-Synchrotron and Humboldt University, Los Alamos National
Laboratories, the TANGO Consortium of Taiwan, the University of Wisconsin at
Milwaukee, and Lawrence Berkeley National Laboratories. Operations are
conducted by COO, IPAC, and UW.

\software{
          \texttt{astropy} \citep{Astropy-Collaboration13},
          \texttt{scipy} \citep{2020SciPy-NMeth}, 
          \texttt{matplotlib} \citep{Hunter07},
          \texttt{pandas} \citep{McKinney10},
          \texttt{emcee} \citep{Foreman-Mackey13},
          \texttt{corner} \citep{Foreman-Mackey16},
          \texttt{SALT2} \citep{Guy07},
          \texttt{sncosmo} \citep{Barbary16},
          \texttt{statsmodels} \citep{Seabold10}
          }


\appendix

\section{Updated Priors Following the Change of Variables}\label{sec:prior}

As mentioned in \S\ref{sec:model}, there is a strong degeneracy in the
posterior estimates of $A$ and $\alpha$. This degeneracy can be removed under
the change of variables from ($A, \alpha$) to ($A^\prime, \alpha^\prime$),
where $A^\prime = A 10^\alpha$ and $\alpha^\prime = \alpha$. From the Jacobian
of this transformation, we find
$$P(A^\prime, \alpha^\prime) = 10^{-\alpha^\prime} P(A,\alpha).$$
The change in variables should not affect the prior probability, therefore
\begin{equation} 
    P(A^\prime, \alpha^\prime) = 10^{-\alpha^\prime} P(A^\prime
10^{-\alpha^\prime},\alpha^\prime), 
\label{eqn:change_prior} 
\end{equation}
which can be satisfied by:
\begin{equation}
    P(A^\prime, \alpha^\prime) \propto {A^\prime}^{-1} 10^{-\alpha^\prime}.
\label{eqn:transform}
\end{equation}
While Equation~\ref{eqn:change_prior} is also satisfied by $P(A^\prime,
\alpha^\prime) \propto {A^\prime}^{-1}$, adopting this as the joint prior on
($A^\prime, \alpha^\prime$) does not remove the degeneracy between the
parameters as $A^\prime$ absorbs the multiplicative factor of $10^\alpha$,
effectively reducing the problem to be the same as it was before the change of
variables. Thus, as listed in Table~\ref{tab:priors}, we adopt
Equation~\ref{eqn:transform} as the prior on the transformed variables, which
we find breaks the degeneracy (see
Figures~\ref{fig:corner_good}--\ref{fig:corner_bad}).

\section{Quality Assurance}\label{sec:qa}

\begin{figure}
    \centering
    \includegraphics[width=3.4in]{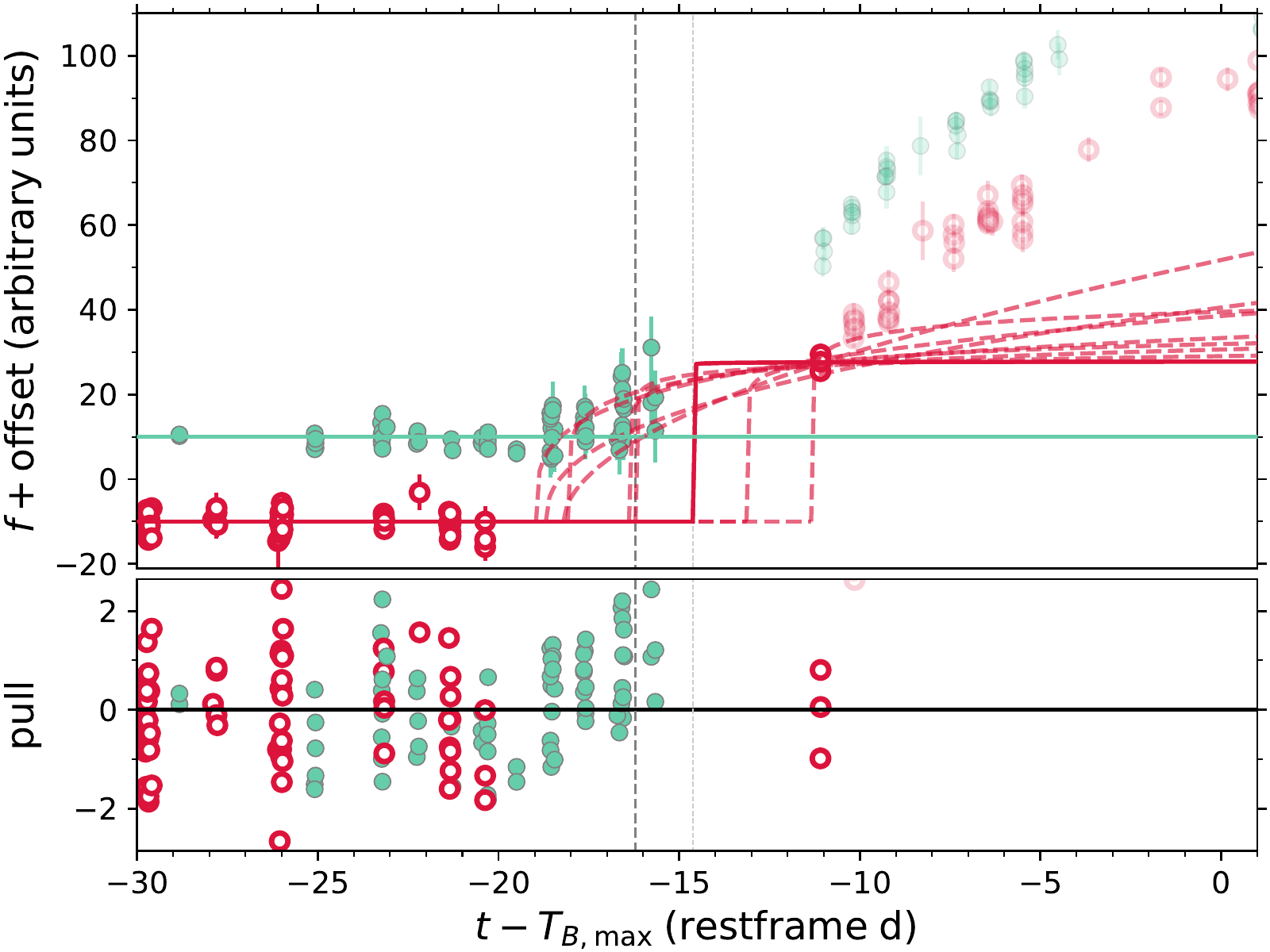}
    \caption{Same as the bottom panel of Figure~\ref{fig:corner_good} for
    ZTF18aaqffyp (SN\,2018bhr), an SN with observations that place very weak
    constraints on \tfl. Marginalized posteriors for $A^\prime$ and
    $\alpha$ are essentially identical to the priors. Posteriors with little
    information beyond the prior are typical of SNe with significant
    observational gaps.}
    \label{fig:biggap_lc}
\end{figure}

As noted in \S\ref{sec:model}, the MCMC model converges for all but one ZTF
SNe within the sample. However, visual inspection of both the corner plots and
individual draws from the posterior quickly reveals that, for some SNe, the
data do not provide strong constraints on the model parameters (see
Figure~\ref{fig:corner_bad}). In the most extreme cases, as shown in
Figure~\ref{fig:biggap_lc}, large gaps in the observations make it nearly
impossible to constrain the model parameters. For these cases, the model
posteriors are essentially identical to the priors (there is always a weak
constraint on \tfl\ from epochs where the SN is not detected).

To identify SNe with poor observational coverage, or unusual structure in the
posterior, we visually examine the light curves and corner plots for each of
the 127 SNe in our sample. We flag SNe where the model significantly
underestimates the flux near \tbmax\ (similar to what is shown in
Figure~\ref{fig:biggap_lc}), as this is a good indicator that the model has
poor predictive value. By definition, the light curve derivative is zero at
maximum light, and the relative change in brightness constantly slows down in
the week leading up to maximum light. Therefore, models of the early
emission should greatly overpredict the flux at maximum, which is why we
adopt this criterion for flagging SNe with poorly constrained model
parameters.

\begin{figure}
    \centering
    \includegraphics[width=3.5in]{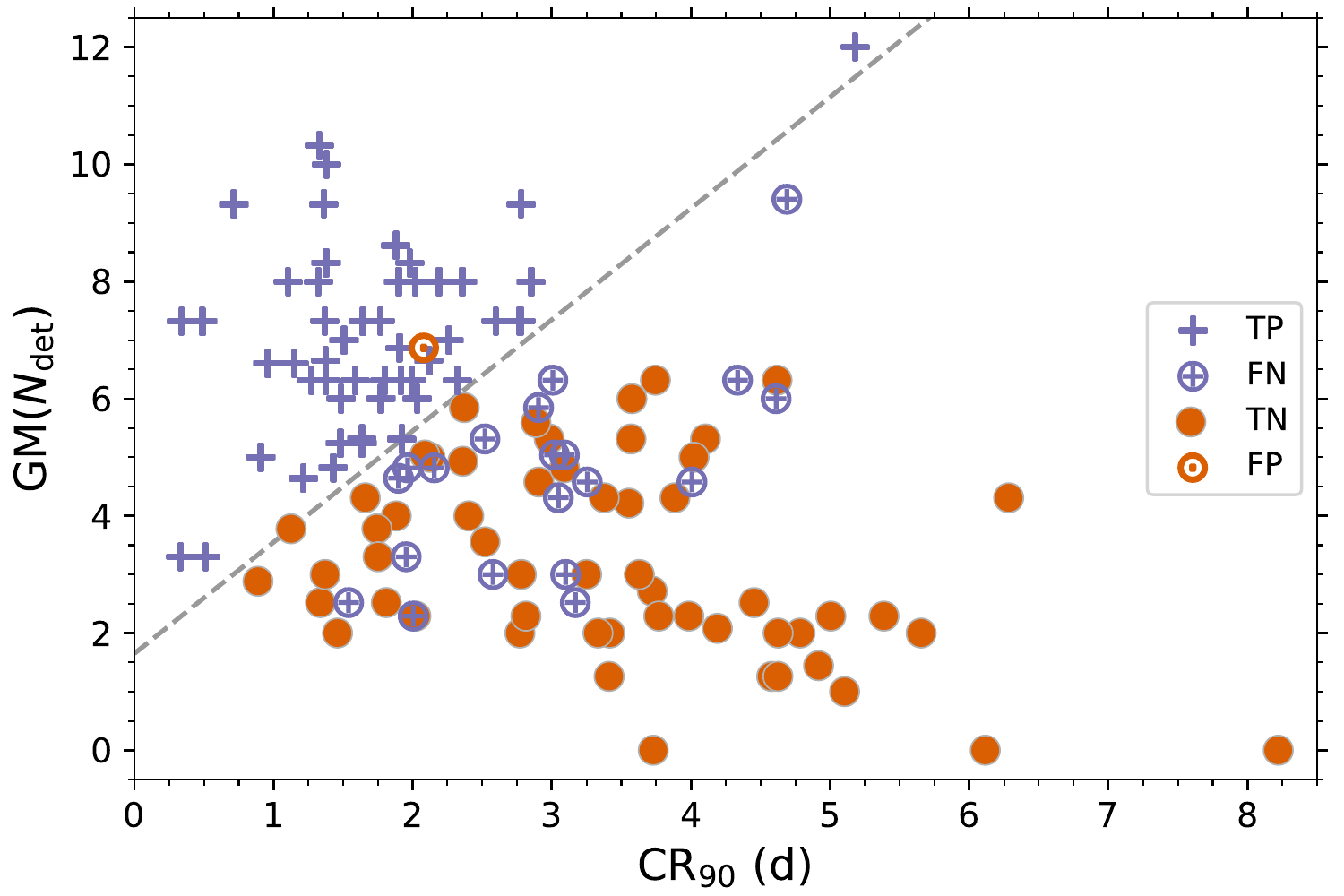}
    \caption{Scatter plot showing the distribution of the 127 ZTF SNe\,Ia in
    the $\mathrm{GM}(N_\mathrm{det})$--$\mathrm{CR}_{90}$ plane. Models with
    flagged posterior parameters are shown as orange circles, while those
    that are not flagged are shown via $+$ symbols. The dashed line shows the
    adopted separation threshold for identifying reliable model fits (above
    the line), and unreliable model fits. FP and FN (see text) SNe are
    circled. }
    \label{fig:flagged_sn}
\end{figure}

Numerically, the visually flagged SNe can, for the most part, be identified
by a combination of two criteria: the 90\% credible region on \tfl,
$\mathrm{CR}_{90}$, and the number of nights on which the SN was detected.
Rather than providing a threshold for detection (e.g., $3\sigma$, $5\sigma$,
etc.), we count all nights with $f_\mathrm{mean} \le 0.4 f_\mathrm{max}$
after the median marginalized posterior value of \tfl\ with observations in
either the \gztf, \rztf, or both filters, $N_{g, \mathrm{det}}$, $N_{r,
\mathrm{det}}$, and $N_{gr, \mathrm{det}}$, respectively. We take the
geometric mean of these three numbers to derive the ``average'' number of
nights on which the SN was detected, $\mathrm{GM}(N_\mathrm{det})$. A scatter
plot showing $\mathrm{GM}(N_\mathrm{det})$ vs.\ $\mathrm{CR}_{90}$ is shown
in Figure~\ref{fig:flagged_sn}. Visually flagged SNe are shown as orange
circles, while $+$ symbols show those that were not flagged.

The visual inspection procedure described above is not fully reproducible
(visual inspection is, by its very nature, subjective). Therefore, we aim to
separate the SNe into two classes (reliable and unreliable) via an automated,
systematic procedure. Treating the visually flagged sources as the negative
class, we regard false positives (i.e., visually flagged SNe that are included
in the final population analysis) to be particularly harmful. Therefore, we adopt
$$ \mathrm{GM}(N_\mathrm{det}) \ge 1.9\,\mathrm{CR}_{90} + 1.65,$$
as the classification threshold for reliable model fits (as shown via the
dashed line in Figure~\ref{fig:flagged_sn}). This threshold retains 50 true
positives (TP; visually good models included in the final sample) with only a
single false positive (FP; visually flagged SNe in the final sample). This
choice does result in 20 false negatives (FN; visually good models
\textit{excluded} from the final sample), while all remaining flagged SNe are
true negatives (TN). Further scrutiny of the FN reveals several light curves
with significant observational gaps--which, as discussed above, makes it
difficult to place strong constraints on the model parameters. Ultimately, our
two-step procedure identifies 51 SNe as reliable, while 76 are excluded from
the final population analysis due to their unreliable constraints on the model
parameters.

\section{Rare and Unusual Thermonuclear SNe}\label{sec:rare}

In \citet{Yao19}, we identified six peculiar SNe\,Ia, which were
classified as either SN\,2002cx-like (hereafter 02cx-like or SN\,Iax),
super-Chandrasekhar (SC) explosions, or SNe\,Ia interacting with their
circumstellar medium (CSM), known as SN Ia-CSM. For this study, we have also
excluded ZTF18abdmgab (SN\,2018lph), a 1986G-like SN that would not typically
be included in a sample used for cosmological studies. Here, we summarize the
early evolution of these events.

\begin{figure}
    \centering
    \includegraphics[width=3.4in]{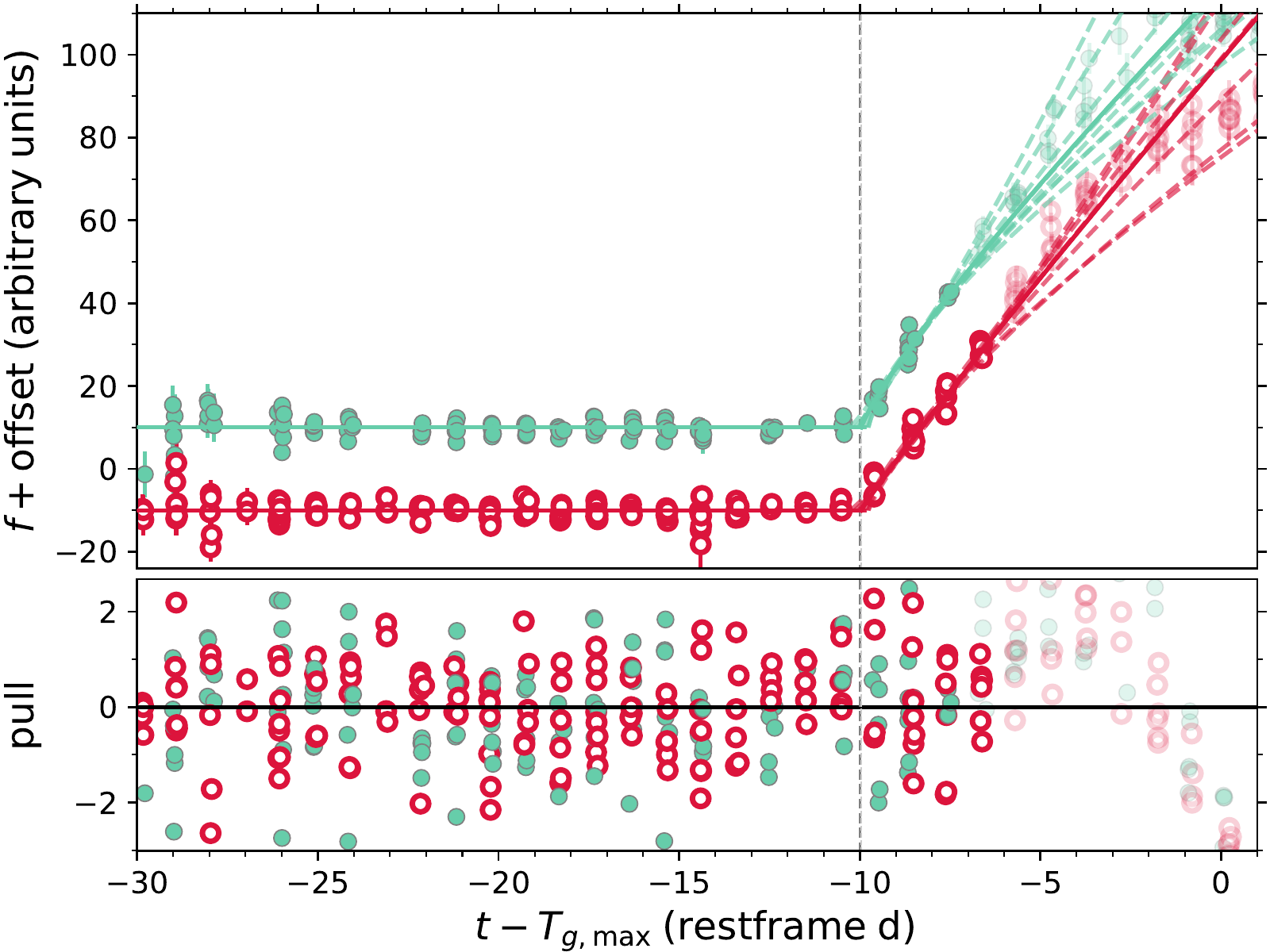}
    \caption{Same as the bottom panel of Figure~\ref{fig:corner_good} for
    ZTF18abclfee (SN\,2018cxk), an 02cx-like SN with strong constraints on
    \tfl, as well as a short rise time ($\sim$10\,d). ZTF18abclfee has the
    tightest constraints on \trise\ of all 02cx-like SNe observed to date. The
    median 1D marginalized posterior value of \tfl\ and the the maximum
    \textit{a posteriori} value of \tfl\ are nearly identical for
    ZTF18abclfee, so the thin gray line showing the latter is not visible.}
    \label{fig:02cx}
\end{figure}

For ZTF18abclfee (SN\,2018cxk), an 02cx-like SN at $z = 0.029$, we
obtained an exquisite sequence of observations in the time before explosion,
as shown in Figure~\ref{fig:02cx}. According to the DIC, $\alpha \ne 2$ is
decisively preferred for this SN. For ZTF18abclfee, we estimate \trise$ =
10.01 \pm^{0.40}_{0.33}$\,d\footnote{Rise times for the unusual SNe discussed
in this appendix are measured relative to $T_{g,\mathrm{max}}$ as
\texttt{SALT2} does not provide reliable estimates of \tbmax\ for non-normal
SNe\,Ia.} (the uncertainties represent the 90\% credible region). This is the
most precise measurement of the rise time of an 02cx-like SN to date. The only
other 02cx-like event with good limits on the rise with deep upper limits is
SN\,2005hk \citep{Phillips07}. SN\,2005hk has a substantially longer rise time
($\sim$15\,d; see \citet{Phillips07}) than ZTF18abclfee, which is not
surprising given that ZTF18abclfee is less luminous and declines more rapidly
than SN\,2005hk \citep{Miller17a,Yao19}. ZTF18abclfee also exhibits a nearly
linear early rise with $\alpha_g = 0.95 \pm^{0.32}_{0.19}$ and $\alpha_r =
0.98 \pm^{0.23}_{0.15}$.

ZTF18aaykjei (SN\,2018crl), a Ia-CSM SN with \trise$ = 22.8
\pm^{2.0}_{1.8}$\,d and $26.3 \pm 1$\,d for the uninformative and $\alpha = 2$
priors, respectively, has a significantly longer rise than the normal SNe in
this study. \citet{Silverman13} points out that Ia-CSM have exceptionally long
rise times, and \citet{Firth15} measure \trise$ > 30$\,d for two of the SNe in
the \citet{Silverman13} sample. We also note that the \rztf\ peak of
ZTF18aaykjei occurs at least one week after the \gztf\ peak, as has been seen
in other Ia-CSM SNe \citep{aldering05gj,prieto05gj}.

There are two SC SNe\,Ia (ZTF18abdpvnd/SN\,2018dvf and
ZTF18abhpgje/SN\,2018eul) and two candidate SC SNe (ZTF18aawpcel/SN\,2018cir
and ZTF18abddmrf/SN\,2018dsx) identified in \citet{Yao19}. Each of these
events exhibits a long rise, $\gtrsim 20$\,d and $\gtrsim 25$\,d for the
uninformative and $\alpha=2$ priors, respectively, as previously seen in other
SC events (e.g., \citealt{Scalzo10,Silverman11}). We note that, with the
exception of ZTF18abdpvnd ($z = 0.05$), these events are detected at high
redshift ($z \gtrsim 0.15$); as a result the constraints on the individual
rise time measurements are relatively weak.

Finally, for ZTF18abdmgab (SN\,2018lph), the 86G-like SN identified in
\citet{Yao19}, we cannot place strong constraints on the rise time, due to a
significant gap in the observations around \tfl.

\section{Systematics}\label{sec:systematics}

\subsection{Definition of ``Early'' for Model Fitting}\label{sec:flux_cut}

In \S\ref{sec:model} we highlighted that there is no single agreed upon
definition of which SN Ia observations are best for modeling the early
evolution of SNe\,Ia. Throughout this study, we have adopted a threshold,
$f_\mathrm{thresh}$, relative to the maximum observed flux, $f_\mathrm{max}$,
whereby we define all observations less than $f_\mathrm{thresh} = 0.4$ the
maximum in each filter ($f_\mathrm{obs} \leq 0.4f_\mathrm{max}$) as the early
portion of the light curve. As noted in \S\ref{sec:model}, setting
$f_\mathrm{thresh} = 0.4$, is arbitrary (although consistent with some
previous studies). Here, we examine the effect of this particular choice if we
had instead adopted $f_\mathrm{thresh} = 0.25$, 0.30, 0.35, 0.45, or 0.50 for
the fitting procedure in \S\ref{sec:model}.

\begin{figure}[ht]
    \centering
    \includegraphics[width=3.4in]{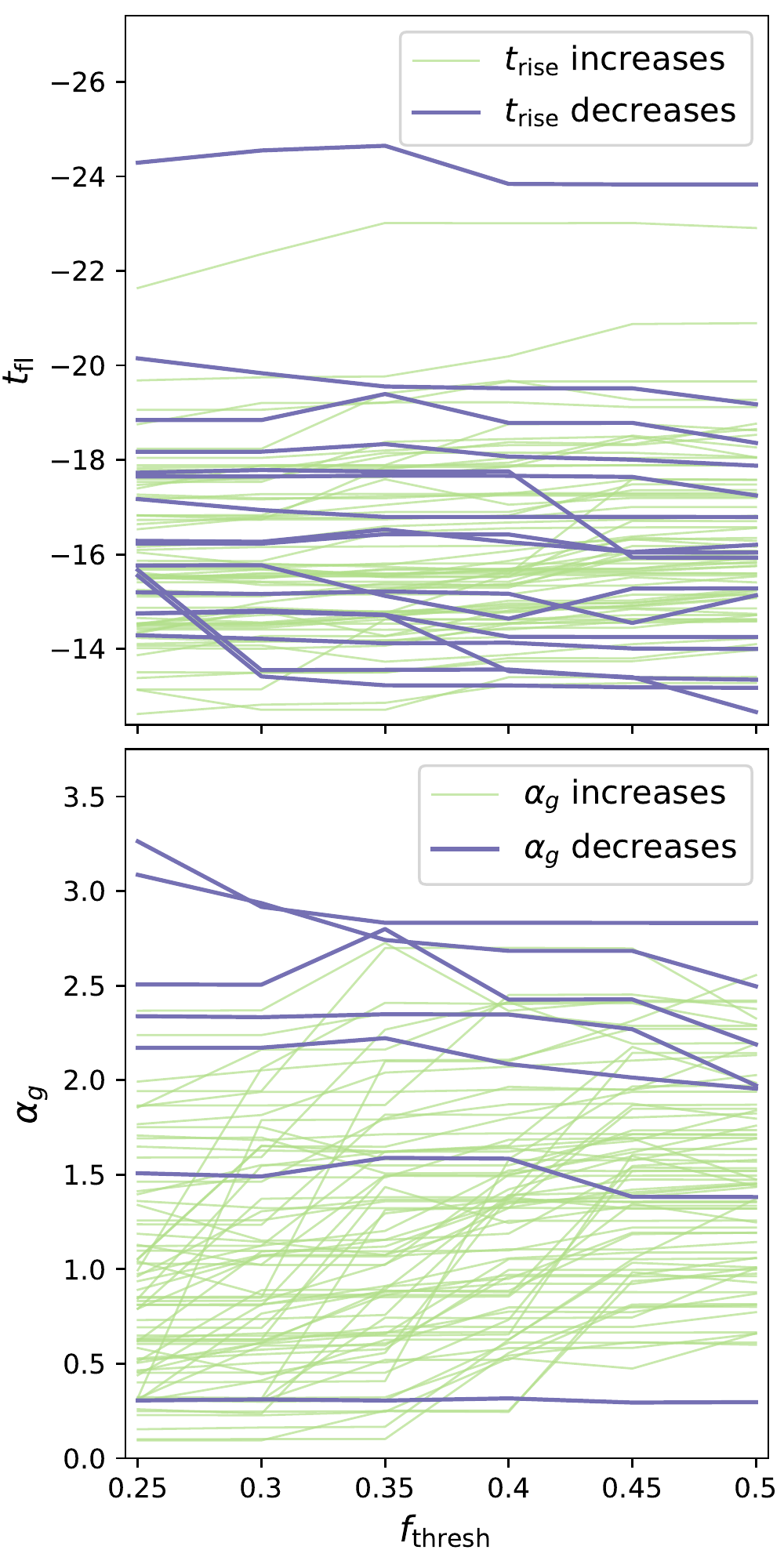}
    \caption{Evolution of the inferred values of \tfl\ and $\alpha_g$ as
    $f_\mathrm{thresh}$ is increased from 0.25 to 0.5. Only SNe with
    consistent model parameters, that nevertheless show evidence for
    increasing or decreasing with $f_\mathrm{thresh}$, are shown (see text for
    a definition of consistent, increasing, and decreasing). Thin green lines
    show SNe where \trise\ or $\alpha_g$ increases as more observations are
    included in the model, while thick purple lines show SNe for which these
    values decline. We find that, for the vast majority of SNe, as additional
    observations are included in the model fit, both \trise\ and $\alpha_g$
    increase.}
    \label{fig:flux_frac}
\end{figure}

There are 12 SNe for which the MCMC chains did not converge for one or more of
the alternative flux thresholds. They are excluded from the analysis below.
For the remaining 115 SNe in our sample, we consider the model parameters to
be consistent if the marginalized, one-dimensional 90\% credible regions for
the three parameters that we care about, \tfl, $\alpha_g$, and $\alpha_r$,
overlap with the estimates when $f_\mathrm{thresh} = 0.40$.\footnote{Given the
strong correlation between $\alpha_g$ and $\alpha_r$ (see
\S\ref{sec:alpha_correlation}), we discuss only $\alpha_g$ below.} This
definition identifies substantial differences in the final model parameters
while varying $f_\mathrm{thresh}$ over a reasonable range. Of the 115 SNe with
converged chains, we find that 98 ($> 85\%$ of the sample) have marginalized,
1D posterior credible regions consistent with the results for
$f_\mathrm{thresh} = 0.40$, independent of the adopted value of
$f_\mathrm{thresh}$. 15 of the 17 SNe that do not have consistent \tfl,
$\alpha_g$, or $\alpha_r$ estimates feature gaps in observational coverage,
which is the likely reason for the inconsistency. As $f_\mathrm{thresh}$
increases from 0.25 to 0.5, the information content dramatically changes
before and after a gap leading to significantly different parameter estimates.

If we alternatively consider the results to be consistent only if the 68\%
credible regions agree with the $f_\mathrm{thresh} = 0.40$ results, then only
64 SNe have consistent parameters as $f_\mathrm{thresh}$ varies. This suggests
that, while the results are largely consistent, the central mass of the
posterior density is affected by which data are included or excluded in the
model fit. In Figure~\ref{fig:flux_frac}, we show how the estimates of \tfl\
and $\alpha_g$ change as a function of $f_\mathrm{thresh}$ for SNe with
consistent model parameters. Note that, by construction, the 90\% credible
regions for each SN overlap at every value of $f_\mathrm{thresh}$, and thus,
for clarity, we omit error bars.

To identify trends with $f_\mathrm{thresh}$, we define SNe with both
$\alpha_g(f_\mathrm{thresh} = 0.5)$ and $\alpha_g(f_\mathrm{thresh} = 0.45)$
greater than both $\alpha_g(f_\mathrm{thresh} = 0.25)$ and
$\alpha_g(f_\mathrm{thresh} = 0.3)$ to show evidence for $\alpha_g$ increasing
with $f_\mathrm{thresh}$. We define $\alpha_g$ as decreasing in cases where
the opposite is true. Of the 98 SNe with consistent model parameters, 75 show
evidence for $\alpha_g$ increasing with $f_\mathrm{thresh}$, while only 7 show
evidence for a decline. Using a similar definition for \trise\ (note that
decreasing \tfl\ corresponds to increasing \trise), we find that in 68 SNe
\trise\ increases with $f_\mathrm{thresh}$, while in 16 SNe \trise\ decreases
as more observations are included in the fit. Thus, the vast majority of SNe
exhibit an increase in $\alpha_g$ and \tfl\ as $f_\mathrm{thresh}$ is
increased. Figure~\ref{fig:flux_frac} shows that the magnitude of this trend
is much larger for $\alpha_g$ than \trise, which makes sense. When there are
few SN detections, which is more likely when $f_\mathrm{thresh}$ is low, small
values of $\alpha$ fit the data well, as in Figure~\ref{fig:biggap_lc}.
Including more information about the rise, by increasing $f_\mathrm{thresh}$,
results in very low values of $\alpha$ no longer being consistent with the
data. On the other hand, \tfl is strongly constrained by the first epoch of
detection (see \S\ref{sec:redshift_correlations}). In this case, the addition
of more observations will not lead to as dramatic an effect.

\subsection{The Importance of Pre-explosion Observations}\label{sec:pre_explosion}

A unique and important component of our ZTF data set is the nightly collection
of multiple observations. \citet{Yao19} demonstrated that such an
observational sequence enables low-S/N detections of the SN prior to the
traditional $5\sigma$ discovery epoch (see \citealt{Masci19}), which can
provide critical constraints on \tfl. Many previous studies have utilized
filtered observations that were obtained $\sim$1\,d or more after the epoch of
discovery (e.g., \citealt{Riess99a,Aldering00,Ganeshalingam11,Zheng17a}). To
demonstrate the importance of the ZTF subthreshold detections, we refit the
model from \S\ref{sec:model} to each of our ZTF light curves after removing
all observations before and on the night the SN is first detected (i.e.,
S/N\,$\geq 5$, as defined in \citet{Yao19}).

Following the removal of these observations, the MCMC chains converge (see
\S\ref{sec:model}) for only 10 SNe. This is understandable as the removal of
the ``baseline'' observations makes it very difficult to constrain $C_d$ and
$\beta_d$. The removal of these observations leads to dramatically different
estimates of the model parameters for these 10 SNe. Thus, we report the
results given the strong trends, though we caution that these results are
somewhat preliminary and should be confirmed with more detailed simulations.

With the baseline observations removed, the inferred value of \tfl\ increases
(i.e., \trise\ decreases) for all 10 SNe relative to the results from
\S\ref{sec:model}. The median difference of this shift is $\sim$3.5\,d. Using
the definition of agreement from Appendix~\ref{sec:flux_cut}, i.e., overlap in
the 90\% credible regions, only three of the ten SNe have estimates of \tfl\
that agree after removing the nondetections. Removing the baseline
observations also decreases estimates of $\alpha$ (which agrees with the trend
seen in Figure~\ref{fig:model_parameters}), with only five of the ten SNe
having estimates of $\alpha_g$ and $\alpha_r$ that agree. These trends suggest
that pre-explosion observations are critically needed to produce accurate
estimates of \trise\ \citep[see also][]{Gonzalez-Gaitan12}.

\bibliography{/Users/adamamiller/Documents/tex_stuff/papers}
\bibliographystyle{aasjournal}


\end{document}